\documentclass[12pt,reqno]{article}
\usepackage{cite,epic,eepic,euscript,verbatim,amsmath,amsthm,amssymb,amsfonts,afterpage,float,bm,stmaryrd,paralist,cancel,color,fancyhdr,graphics,graphicx,epsf,hyperref,makeidx,epsfig}

\usepackage{subfigure,epstopdf}
\usepackage[titletoc]{appendix}

\renewcommand{\theequation}{\thesection.\arabic{equation}}
\numberwithin{equation}{section}
\numberwithin{table}{section}

\newcommand{\reff}[1]{{\rm (\ref{#1})}}

\newcommand{\ve}{\varepsilon}          
\newcommand{\bx}{\bm x}            
\newcommand{\bS}{\bm S}            
\newcommand{\bX}{\bm X}            
\newcommand{\by}{\bm y}            
\newcommand{\bu}{\bm u}

\def\XXint#1#2#3{{\setbox0=\hbox{$#1{#2#3}{\int}$}
\vcenter{\hbox{$#2#3$}}\kern-.51\wd0}}

\begin{document}

\title
{
Energetic Variational Approach for Prediction of Thermal Electrokinetics in Charging and Discharging Processes of Electrical Double Layer Capacitors
}
\author{
Xiang Ji
\thanks{Department of Mathematics,
	Soochow University, 1 Shizi Street, Suzhou 215006, Jiangsu, China.
	Email: xji@stu.suda.edu.cn.}
\and
Chun Liu
\thanks{
	Department of Applied Mathematics, Illinois Institute of Technology, Chicago, IL 60616, USA. Email: cliu124@iit.edu.	
}
\and
Pei Liu
\thanks{
	School of Mathematics, University of Minnesota, Twin Cities, MN 55455, USA.  Email: liu01304@umn.edu.
}
\and
Shenggao Zhou
\thanks{School of Mathematical Sciences, MOE-LSC, and CMA-Shanghai, Shanghai Jiao Tong University, Shanghai 200240, China. Email: sgzhou@sjtu.edu.cn. }
}

\date{\today}

\maketitle
\begin{abstract}
This work proposes a new variational, thermodynamically consistent model to predict thermal electrokinetics in electric double layer capacitors (EDLCs) by using an energetic variational approach. The least action principle and maximum dissipation principle from the non-equilibrium thermodynamics are employed to develop modified Nernst-Planck equations for non-isothermal ion transport with temperature inhomogeneity.  Laws of thermodynamics are employed to derive a temperature evolution equation with heat sources due to thermal pressure and electrostatic interactions.  Numerical simulations successfully predict temperature oscillation in the charging-discharging processes of EDLCs, indicating that the developed model is able to capture reversible and irreversible heat generations. The impact of ionic sizes and scan rate of surface potential on ion transport, heat generation, and charge current is systematically assessed in cyclic voltammetry simulations.  It is found that the thermal electrokinetics in EDLCs cannot follow the surface potential with fast scan rates, showing delayed dynamics with hysteresis diagrams. Our work thus provides a useful tool for physics-based prediction of thermal electrokinetics in EDLCs.

\bigskip

\end{abstract}

{ \allowdisplaybreaks
\section{Introduction}
\label{s:Introduction}
Immense demand for green energy has greatly promoted the development of electrical energy storage technologies. 
Electric double layer capacitors (EDLCs), also known as supercapacitors, have received significant attention in recent years for their unique features and broad application spectrum~\cite{FullerHarb_2018Book}. Electric energy is stored in electric double layers forming at solid/liquid interface. In contrast to traditional dielectric capacitors, EDLCs can achieve much larger energy densities due to small charge separation in nanoscale electric double layers~\cite{Conway::1999}. In contrast to conventional rechargeable batteries, EDLCs have higher charging/discharging efficiency, longer cycle life, and higher power density~\cite{Xu_nature14, yang_sci13, Schiffer_JPS06,Ramon_JCIS18,Burke00}. 


The heat generation is a major concern for the function of EDLCs in that they are often cycled under high current density, which may result in an unexpected local temperature rise. In practice, excessive temperature rise in EDLCs could cause various irreversible damages, e.g., accelerated aging~\cite{Miller_ea_06}, enhanced self-discharge rates~\cite{Xiong2015, Sakka_jps_09, GUILLEMET_JPS06},  capacitance reduction~\cite{Torregrossa_16, Torregrossa_16_2}, and electrolytes decomposition/evaporation~\cite{Masarapu_acs_09}. To avoid these issues,  temperature elevation in EDLCs during operation should be suppressed. Therefore, comprehensive understanding on heat generation is of significance and highly desirable  for the design, optimization, and manufacture of new generation EDLCs, to ensure the safety and long-term stability. 

Temperature oscillations observed in the charging and discharging experiments of EDLCs evidence the contribution of the reversible and irreversible heat generation~\cite{Schiffer_JPS06, Pascot_TA10, Dandeville_TA11, Zhang_TA16}.  Many existing models on heat generation neglect the underlying thermal electrokinetic details in the operation of EDLCs. Either uniform heat generation for the whole device is assumed~\cite{Gualous_IEEE00, Sakka_jps_09, Dandeville_TA11, Gualous_IEEE11} or only irreversible heat generation is accounted for with parameters determined by experiments~\cite{Schiffer_JPS06, Sakka_jps_09, Sauer_JPS07}.   The irreversible Joule heating effect, which is proportional to the square of charge current, often accounts partially for the temperature change in EDLCs during charging and discharging processes~\cite{Porras_ep03, Pilon_JPS15, Janssen_PRL17}. Without reversible heat generation, theoretical models fail in predicting the temperature recession in the discharging process. 

Much attention has been paid to the first-principle modeling of reversible and irreversible heat generation in EDLCs in recent years. 
Based on a thermodynamic viewpoint, Schiffer \emph{et al.} \cite{Schiffer_JPS06} took into account the reversible heat generation rate as $-T\frac{d S}{d t}$, where the entropy change of ions due to the formation and dissolution of electric double layers is estimated by using the volume of Stern layers.  The derived model can successfully predict the temperature oscillation in the charging-discharging processes.  
Zhang \emph{et al.} proposed to describe the reversible heat via the relationship between surface Gibbs free energy and electric energy at the electric double layer interface~\cite{Zhang_TA16}. Simulation results indicated that the proposed model could capture temperature oscillation during charging and discharging cycles and very well fit the experimental temperature profiles.
To get physically faithful models with spatiotemporal resolution, d'Entremont~\emph{et al.}~\cite{pilon_jps14} developed an energy conservation equation for temperature, coupling electrodiffusion, heat generation, and thermal transport. The model incorporates irreversible Joule heating and reversible heat generation due to electrodiffusion, steric effect, and entropy changes. It is further extended to consider the effect of asymmetric ionic sizes on thermal and charge dynamics~\cite{Pilon_JPS15}.  Numerical simulations demonstrated that the model is capable of predicting the temperature oscillation in charging-discharging processes. To reduce computational cost, scaling law analysis was performed to gain insights on reversible heating and develop a guideline for thermal management strategies~\cite{Plion_IJHMT_14}.

Another type of model is derived by considering a source term that is the dot product of current and the electric field, including irreversible and reversible heat generation~\cite{Janssen_PRL17}. It is demonstrated that in terms of predicting the temperature change, the model has a good agreement with a thermodynamic identity involving temperature and electrostatics in the slow charging process. More recently, reversible heat production is studied based on a model that is derived by considering the ratio of reversible heat production into EDLCs during an isothermal charging process and the electric work~\cite{Janssen_jcp_21}. It should be pointed out that the abovementioned models ignore porous structure of electrodes and treat EDLCs as a simple one-dimensional system with planar electrodes. Recent years have seen the development of models to evaluate the impact of cylindrical or porous electrodes on the performance of EDLCs~\cite{GirardPilon_EA16, MeiPilon_EA17, LianJanssen_PRL20, Kundu_jps_21, YangJanssenLianRoij_jcp_22, LinLian_PRL22, HuangLianLiu_AJ22}. For instance, a stack-electrode model with an electric circuit analogy was proposed to mimic the porous electrodes, and applied the model to study the relation between relaxation timescale and porosity~\cite{LianJanssen_PRL20, LinLian_PRL22}. The model was also extended to investigate the influence of structural parameters of porous electrodes on the temperature rise during the charging process of EDLCs~\cite{HuangLianLiu_AJ22}.

Despite irreversible and reversible heat generations have been included, the abovementioned models describe ion transport by modified Nernst-Planck (NP) equations with ionic size effects~\cite{Bazant_PRE07II,  Zhou_pre_11, LiLiuXuZhou_Nonliearity13, Pilon_JPS15, JiZhou_CMS19}, which ignore the impact of inhomogeneous temperature distribution on the ion transport as in isothermal scenarios. The effect of spatio-temporal inhomogeneity in temperature on charge dynamics is expected to become significant as large temperature rises in EDLCs~\cite{pilon_jps14}. Due to the multi-scale and multi-physics nature, the development of thermodynamically consistent models for the charging/discharging processes of EDLCs has been a challenging issue.

In this work, we propose a variational,  thermodynamically consistent model to predict thermal electrokinetics in EDLCs by using an energetic variational approach, which has been proved to be a powerful tool in studying numerous complex multi-scale, multi-physics systems~\cite{Liu_book}, such as liquid crystals~\cite{Sun_DCDS_09}, multiple-phase flows~\cite{Liu_03}, and ionic solutions~\cite{BobHyonLiu_JCP10}. The least action principle and maximum dissipation principle from the non-equilibrium thermodynamics are employed to derive modified NP equations that take into account the diffusion, ionic steric effect, and convection due to the gradient of temperature and electrostatic potential. Laws of thermodynamics are employed to derive temperature evolution equations with heat sources arising from thermal pressure and electrostatic interactions. Extensive numerical simulations based on the proposed model are performed to understand the  thermal electrokinetics in the charging-discharging processes of EDLCs.

\section{Methods}
\label{s:Model}

\begin{figure}[bhtp]
	\centering
	\includegraphics[scale=.32]{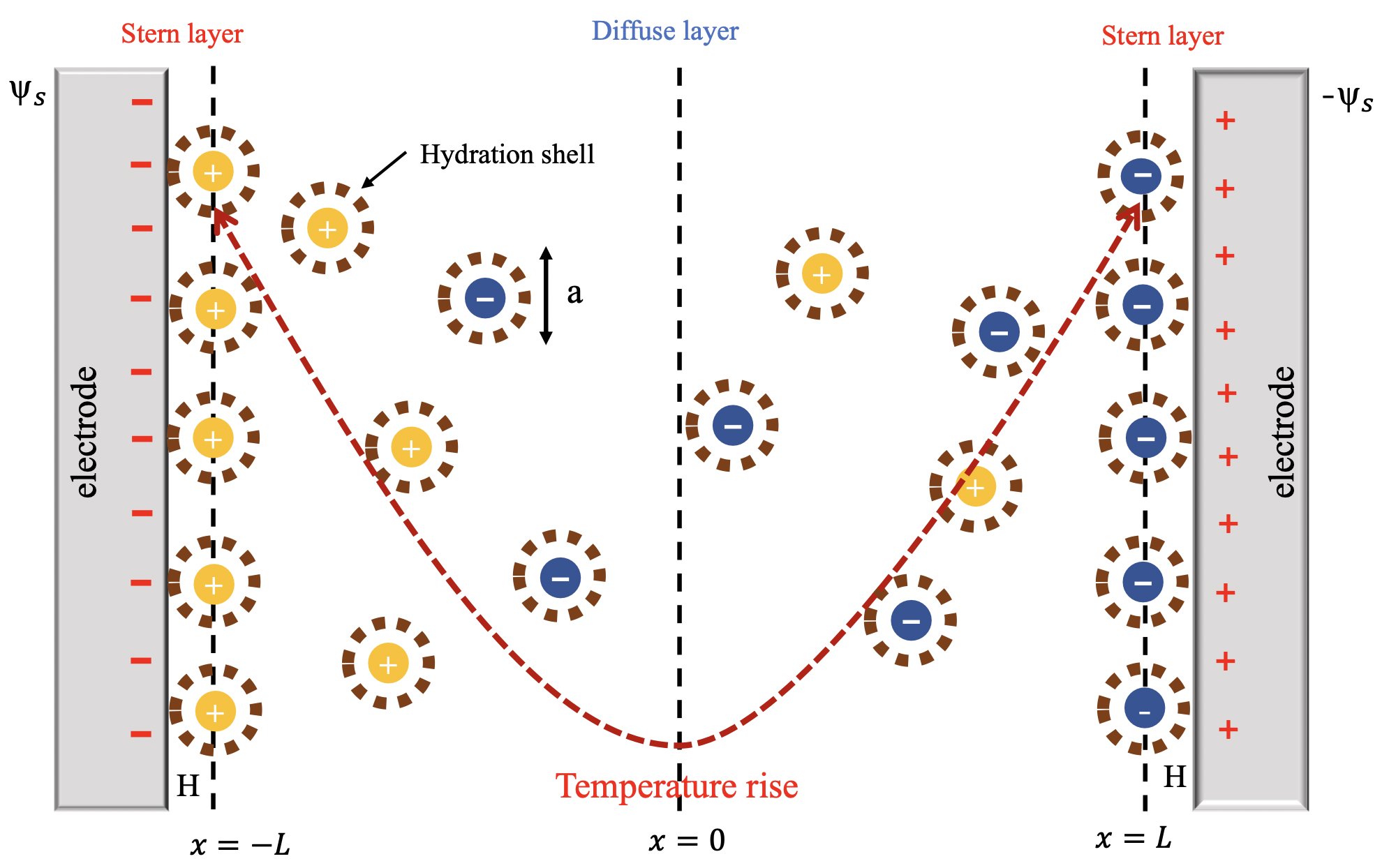}
	\caption{ A schematic view of an electric double layer capacitor. Hydrated cations and anions are assumed to have a uniform diameter $a$. Biased potential differences are applied across two thermally-insulated,  blocking parallel electrodes that are located at $x = -L-H$ and $x = L+H$ with $H$ being the width of Stern layers.}
	\label{f:geometry}
\end{figure}
Consider a one-dimensional electric double layer capacitor (EDLC) that consists of binary, symmetric electrolytes blocked by two parallel electrodes located at $x = -L-H$ and $x = L+H$; cf. a schematic plot of the system shown in Fig.~\ref{f:geometry}. Hydrated cations and anions are assumed to have a uniform diameter $a$. With applied biased potential difference exerted by the electrodes, electric double layers due to electrostatic interactions are formed at electrodes. In electrolytes, a Stern layer right adjacent to each electrode is assumed to have width $H=a/2$, and it is followed by a diffuse layer. It is assumed that there is no electrochemical reaction at the interface between electrolytes and electrodes. 

\subsection{Energetic variational approach}
We now propose an energetic variational model for the description of ionic electrodiffusion, heat generation, and thermal transport in EDLCs. Although a one-dimensional EDLC is mainly studied in present work, the model derived here is valid for general three dimensional cases.
 
To characterize the charging and discharging processes, we denote by $T(\bx,t)$, $\psi(\bx,t)$, and $\rho_i(\bx,t)$ the temperature distribution, electric potential, and ionic concentration at location $\bx$ for time $t$, respectively. Let $\Omega $ be the domain for an EDLC system under consideration.  For any arbitrary subdomain $V\subset \Omega$, the mean-field electrostatic free-energy functional $F(V, t)$ is a functional of the particle densities and temperature given by
\begin{align}
	\label{E:freeE}
	F(V,t) = F_{pot}(V, t) + F_{ent}(V, t).
\end{align}
Here the electrostatic potential energy reads
\begin{equation}\label{ElePot}
	\begin{aligned}
	F_{pot}(V,t) &= \sum_{i,m = \pm}\frac{q_iq_m}{2}\iint_{V}\rho_i(\bx,t)\rho_m(\bx^{\prime},t)G(\bx,\bx^{\prime})d\bx d\bx^{\prime} \\   
	&\qquad + \sum_{i= \pm} q_i\int_{V}\rho_i(\bx)\left( \psi_X(\bx,t) + \int_{\Omega \backslash V} \sum_{m = \pm} \rho_m(\bx^{\prime}, t)G(\bx,\bx^{\prime})d\bx^{\prime}\right) d\bx,
\end{aligned}
\end{equation}
where $q_i= z_i e$ with $z_i$ being the ionic valence. The first term in $F_{pot}(V,t)$ represents for the electrostatic potential energy inside $V$ from the Coulomb interactions, with the Green function $G$ satisfying $-\nabla\cdot\epsilon_r\epsilon_0\nabla G(\bx,\bx^{\prime}) = \delta(\bx, \bx^{\prime})$. Here $\epsilon_0$ is the absolute permittivity and $\epsilon_r$ is the dielectric coefficient.  The second term is the electrostatic potential energy from the external fields, including contributions from ions outside $V$ and an external electric potential $\psi_{X}$, e.g., the boundary potential on the electrodes.
The entropy contribution in \reff{E:freeE} is given by
\begin{equation}
	\begin{aligned}
	\label{ent_origin}
	F_{ent}(V,t)  &= \sum_{i = 0, \pm}\int_{V} k_B T(\bx,t)\rho_i(\bx,t)[\log(v\rho_i(\bx,t))-C_i\log T(\bx,t)]d\bx,
	\end{aligned}
\end{equation}
where $k_B$ is the Boltzmann constant and $C_i$ is a constant related to the heat capacitance of each species. Note that the subscript $i=0$ stands for the solvent molecules, and uniform size of both ions and solvent molecules, denoted by $v=a^3$, is assumed here. The steric effect is taken into account via the inclusion of solvent entropy, which is described by a lattice-gas model for uniform-sized solvent molecules and ions~\cite{BAO_PRL97, Bazant_PRE07I}. An upper limit $1/v$ is imposed on the local density via the constraint 
\begin{align}
	\sum_{i = 0,\pm}v\rho_{i}(\bx,t) = 1.
\end{align}
For simplicity, we assume that $C_+ = C_- = C_0$ for each species. As such, the entropy contribution reads
\[
	\begin{aligned}
	F_{ent}(V,t) &= \int_{V} \Psi\left( \rho_+(\bx,t),\rho_-(\bx,t), T(\bx,t) \right)  d\bx  \\
			&:= \int_{V} \left[ \sum_{i = \pm}\Psi_i\left(\rho_{i}(\bx, t), T\left(\bx, t \right)  \right)  + \Psi_0\left( \rho_{+}(\bx,t), \rho_{-}(\bx, t), T(\bx, t)\right) -\Psi_T(T(\bx, t))\right] d\bx,
	\end{aligned}
\]
where $\Psi_{\pm}, \Psi_0$, and $\Psi_T$ are defined as
\begin{equation}\label{ent_re}
		\begin{aligned}
			\Psi_{\pm}\left(\rho_{\pm}(\bx, t), T\left(\bx, t \right)  \right)  &:= k_BT(\bx,t) \rho_{\pm}(\bx,t)\log\left( v\rho_{\pm}(\bx,t)\right) ,\\
			\Psi_0\left(\rho_{+}(\bx, t), \rho_{-}(\bx, t), T\left(\bx, t \right)  \right) &:= k_BT(\bx,t)\frac{1-\sum\limits_{i = \pm}v\rho_i(\bx,t)}{v}\log(1-\sum_{i = \pm}v\rho_i(\bx,t)),\\
			\Psi_T\left( T(\bx, t)\right)  &:= \frac{k_BC_0}{v}T(\bx,t)\log T(\bx,t).
		\end{aligned}
\end{equation}
Then the entropy is given by
\begin{align}
\label{S}
S(V,t) = -\int_{\Omega}\frac{\partial\Psi\left( \rho_+(\bx,t),\rho_-(\bx,t), T(\bx,t) \right)  }{\partial T(\bx,t)}d\bx.
\end{align}
Application of the Legendre transform of the free energy leads to the internal energy
\begin{align}
	\label{U}
	U(V,t) = F(V,t) -\int_{\Omega}T(\bx,t)\frac{\partial \Psi\left( \rho_+(\bx,t),\rho_-(\bx,t), T(\bx,t) \right)  }{\partial T(\bx,t)}d\bx.
\end{align}

To characterize the thermal electrokinetics in EDLCs, we propose an energetic variational model that combines the non-equilibrium statistical mechanics and nonlinear thermodynamics~\cite{XuShengLiu_CMS14, BobHyonLiu_JCP10, LiuWuLiu_CMS18}. Central in this model is the energy dissipation law
\begin{align}
	\label{EvaDL}
	\frac{d E^{\rm tot}}{d t}=-\Delta,
\end{align}
where $E^{\rm tot}$ is the total energy and the dissipation functional $\Delta$ often is a linear combination of the squares of various rate functions,  such as velocity and rate of strain. In the energetic variational approach, the Least Action Principle (LAP) is used to derive conservative forces and the Maximum Dissipation Principle (MDP) is employed to obtain dissipative forces. The balance of total forces gives the governing equations for the motion of ions. 

We first introduce the Lagrangian and Eulerian coordinate systems. Let $\Omega_0^X$ be the reference configuration  and $\Omega_t^x$ be the deformed configuration of ions. Denote by $\bX \in \Omega_0^X$ the Lagrangian coordinate and $\bx \in \Omega_t^x$ the Eulerian coordinate. Introduce the flow maps that are defined by 
\begin{equation}\label{FlowMap}
\frac{\partial }{\partial t} \bx_{\pm}(\bX, t) = \bu_{\pm},~ t> 0 ~ \mbox{and}~ \bx(\bX , 0) =\bX,
\end{equation}
where $\bu_{\pm}$  are velocities fields for cations and anions. By mass conservation, the ionic densities satisfy the kinematic equations
\begin{align}\label{masscon}
	\frac{\partial \rho_{\pm}}{\partial t} + \nabla \cdot(\rho_{\pm} \bm{u}_{\pm}) = 0.
\end{align}
Alternatively, the kinematic mass conservation equations in the Lagrangian coordinate read~\cite{Forster_13}
\begin{align}
	\label{massconL}
	\rho_{\pm}\left( \bx_{\pm}(\bX, t), t\right) = \frac{\rho_{\pm}(\bX, 0)}{\det \mathcal{F}}, 
\end{align}
where  $\mathcal{F} (\bX,t) =  \frac{\partial \bx(\bX,t)}{\partial \bX}$ is the deformation gradient tensor.
 
The Least Action Principle states that the variation of the action functional with respect to the flow map gives the conservative force $\bm{F}_{con, \pm} = \frac{\delta A\left( \bx_{+}(\bX, t),  \bx_{-}(\bX, t)\right) }{\delta \bx_{\pm}}$, where the action functional is defined as 
\[
\begin{aligned}
A\left(\bx_{+}(\bX, t),  \bx_{-}(\bX, t)\right) &= -\int_{0}^{t^*}  F_{pot}(\Omega, t) + F_{ent}(\Omega, t) dt \\
&= A_{pot}\left( \bx_+(\bX,t),  \bx_{-}(\bX, t)\right) +A_{ent}\left( \bx_+(\bX,t),  \bx_{-}(\bX, t)\right)
\end{aligned}
\]
for some time $t^*>0$. Here the kinetic energy of particles is assumed to be negligible. 

Taking the derivation of conservative forces for cations as an example, we obtain by taking variation of  actions with respect to the flow map $\bx_{+}(\bX, t)$:
\begin{align}
	\bm{F}_{con, +}^{ent} &= \frac{\delta A_{ent}\left( \bx_+(\bX,t),  \bx_{-}(\bX, t) \right) }{\delta \bx_+(\bX,t)} \notag\\
	&= -\left[k_BT\nabla \rho_+ + k_BT\frac{v\rho_+}{1-\sum\limits_{i = \pm}v\rho_i}\sum_{j = \pm}\nabla \rho_j  +\frac{\partial P_+}{\partial T}\nabla T\right], \label{F^ent_con+}
\end{align}
where the thermal pressure for cations is given by
\[
P_+ = k_BT\left(\rho_{-}-\frac{1}{v}\right)\log\left(1-\sum\limits_{j = \pm}v\rho_j\right).
\]
The mathematical details on the derivation of \reff{F^ent_con+} are presented in the \emph{Supporting Information}.
The conservative force due to the electrostatic potential energy can be analogously derived as follows~\cite{XuShengLiu_CMS14, BobHyonLiu_JCP10}:
\begin{align}
         \bm{F}_{con, +}^{pot}&=\frac{\delta A_{pot}\left( \bx_+(\bX,t),  \bx_{-}(\bX, t) \right)}{\delta \bx_+} \notag \\
		           &= \frac{\delta}{\delta \bx_+}\int_{0}^{t^*}-\left[  \sum_{i,m = \pm}\frac{q_iq_m}{2}\iint_{\Omega_t^x}\rho_i(\bx,t)\rho_m(\bx^{\prime},t)G(\bx,\bx^{\prime})d\bx d\bx^{\prime} + \sum_{i= \pm} q_i\int_{\Omega_t^x}\rho_i(\bx)\psi_X(\bx,t)  d\bx  \right] dt  \notag \\  
	&= -q_+\rho_{+}\nabla\phi,  \label{F^pot_con+}
\end{align}
where 
$\phi(\bx, t) = \psi_X(\bx, t) + \sum\limits_{i = \pm}q_i\int_{\Omega}\rho_{i}(\bx^{\prime}, t) G(\bx,\bx^{\prime}) d\bx^{\prime}$ is the electric potential.  Combination of $\bm{F}_{con, \pm}^{ent}$ and $\bm{F}_{con, \pm}^{pot}$ leads to the conservative force for cations and anions:
\begin{align}\label{con}
	\bm{F}_{con, \pm} = -\left( k_BT\nabla \rho_\pm + k_BT\frac{v\rho_\pm}{1-\sum_{i = \pm}v\rho_i}\sum_{j = \pm}\nabla \rho_j+q_\pm \rho_\pm \nabla \phi +\frac{\partial P_\pm}{\partial T}\nabla T \right).
\end{align}

In order to predict the irreversible heat generation, we consider  entropy production functional $\Delta(V, t) = \int_V\widetilde\Delta(\bx,t)d\bx$ for an arbitrary domain $V$, with the entropy production density given by
\begin{align}
\label{delta}
\widetilde{\Delta} = \sum_{i = \pm}\frac{\nu_i\rho_i\vert \bu_i\vert^2}{T}+\frac{1}{k}\vert \frac{\bm{j}_h}{T}\vert^2.
\end{align} 
Here $\nu_i$ is the viscosity of ions of the $i$th species, $\bm{j}_h$ represents for the heat flux, and $k$ is a constant related to the heat conductance. We employ the Maximum Dissipation Principle (MDP) to derive dissipative forces by taking variation of the dissipation functional $\Delta$ with respect to the rates (velocity) in Eulerian coordinates, i.e.,
\begin{align}
\label{diss}
	\bm{F}_{dis, +}(\bx,t) = \frac{T}{2}\frac{\delta \Delta(\Omega,t)}{\delta \bu_{+}(\bx,t)} = \nu_{+}\rho_{+}\bu_{+}.  
\end{align}
Note that the factor $\frac{1}{2}$ is included due to the convention that the energy dissipation is always a quadratic function of certain rates. 
 
Therefore, the balance of conservative forces and dissipative forces yields
\begin{align}
\label{forbal}
\nu_\pm\rho_\pm\bu_\pm = -\left( k_BT\nabla \rho_\pm +\frac{\partial P_\pm}{\partial T}\nabla T+ k_BT\frac{v\rho_\pm}{1-\sum\limits_{i = \pm}v\rho_i}\sum_{j = \pm}\nabla \rho_j+q_\pm\rho_\pm\nabla \phi\right).
\end{align}
Combining the the mass conservation equation~\reff{masscon}, one obtains modified Nernst-Planck (NP) equations that include ionic steric effect and thermal effect for ion transport:
\begin{align}
\label{npp}
\frac{\partial \rho_\pm}{\partial t} = \frac{1}{\nu_\pm}\nabla\cdot\left( k_BT\nabla \rho_\pm +\frac{\partial P_\pm}{\partial T}\nabla T+ k_BT\frac{v\rho_\pm}{1-\sum\limits_{i = \pm}v\rho_i}\sum_{j = \pm}\nabla \rho_j+q_\pm \rho_\pm\nabla \phi\right).
\end{align}
Note that the second term in the right side of the modified NP equation~\reff{npp} can be regarded as the Soret effect that describes ion motion due to temperature gradient~\cite{Majee_pre_11,Stout_pre_17}. In addition, it is remarked that for an isothermal case, the modified NP equations~\reff{npp} reduces to the modified NP equations with ionic steric effects in the work~\cite{Bazant_PRE07I}. 

With~\reff{forbal}, we define the ionic flux density by $\bm{j}_\pm = \rho_\pm\bu_\pm$. Since the electrodes are two blocking walls without electrochemical reaction, zero-flux boundary conditions are imposed at the Stern/diffuse layer interface at $x = -L $ and $x = L$, i.e., 
\begin{align}\label{JBC}
	 \bm{j}_\pm(-L,t) = 0  ~~\mbox{and}~~ \bm{j}_\pm(L,t) = 0.
\end{align}
\subsection{Poisson's equation}
The electric potential $\phi(\bx,t)$ is determined by the Poisson's equation
\begin{align}\label{eq:poi}
-\nabla\cdot\epsilon_0\epsilon_r\nabla\phi (\bx,t)= \sum_{i = \pm}q_i\rho_i (\bx,t).
\end{align}
To take the Stern layers of width $H$ into account, we impose Robin boundary conditions for the electric potential at $x = -L $ and $x = L$ as shown in Fig.~\ref{f:geometry}:
\begin{equation}\label{RBC}
\phi (\pm L,t)\pm H\frac{\partial \phi}{\partial x} (\pm L,t) = \mp\phi_s(t),
\end{equation}
where $\phi_s$ is the applied potential at the electrode surface.  In Cyclic Voltammetry (CV) measurements, $\phi_s(t)$ is prescribed periodically and linearly with time as~\cite{WangPilon_EA12}
\begin{equation}\label{CVpsi}
 \phi_s(t)=\left\{
\begin{aligned}
&\phi_{min}+\nu t,                 & 2(n-1)t_0  \leq t  <(2n-1)t_0,   \\
&\phi_{max}-\nu \left[t-(2n-1)t_0\right],              & (2n-1)t_0 \leq t <2nt_0,
\end{aligned} 
\right. 
\end{equation}
where $\nu$ is the scan rate in $V/s$ and $n (=1,2,3,...)$ is the number of cycles. In such CV measurements, $\tau_{cv} = (\phi_{max}-\phi_{min})/\nu$ denotes the half cycle period.
\subsection{Energy conservation equation}
The first law of thermodynamics that describes energy conservation is employed here to derive the governing equation for temperature evolution. Consider an arbitrary control volume $V$ that does not move along with the flow maps of ions. The rate of work done on $V$ is given by 
\begin{equation}
\begin{aligned}
	\label{dW}
	\frac{d}{dt}W(V,t) &= \sum_{i = \pm}q_i\int_{V}\rho_{i}\frac{\partial}{\partial t}\left[  \psi_X(\bx, t) + \sum_{m = \pm}q_m\int_{\Omega \backslash V}\rho_m(\bx^{\prime})G(\bx,\bx^{\prime})d\bx^{\prime} \right] d\bx  \\ 
	&\quad + \sum_{i = \pm}\int_{\partial V} -P_i(\bx,t)\bu_i(\bx, t)\cdot d\bS,   
\end{aligned}
\end{equation}
where the first term represents the rate of work done by a time-dependent electric potential outside $V$, including the contributions from ions in the domain $\Omega \backslash V$ and external potential $\psi_X$. The second term describes the rate of work done by the thermal pressure $P_i$ through the boundary of $V$. The rate of heat transfer is given by the heat flux $\bm{j}_h$ through $\partial V$:
\begin{align}
	\label{dQ}
	\frac{d}{dt}Q(V,t) = -\int_{\partial V}\bm{j}_h\cdot d\bS.
\end{align}
Since $V$ does not move with the flow maps of ions, we need to consider the total energy flux $J_E$ through the boundary:
\begin{align}
	\label{JE}
	J_E(V,t) = \sum_{i = \pm}\int_{\partial V}\left( q_i\rho_i\phi +e_i^{int}  \right) \bu_i \cdot d\bS,
\end{align}
where $e_i^{int}  = \Psi_{i}-T\frac{\partial \Psi_{i}}{\partial T}$ is the internal energy density, with $\Psi_{i}$ defined in~\reff{ent_re}. Note that $e_{\pm}^{int}=0$ due to the linearity of $\Psi_{\pm}$ with respect to $T$. By the arbitrariness of the control volume $V$, we derive from the energy conservation 
\begin{align}
	\label{th1law}
	\frac{d}{dt}U(V,t) + J_E(V,t) = \frac{d}{dt}W(V,t) +\frac{d}{dt} Q(V,t),
\end{align}
\reff{U}, and \reff{dW}-\reff{JE} that 
\begin{equation}
	\label{th1eq}
	\frac{k_BC_0}{v}\frac{\partial T}{\partial t}  = \nabla\cdot (k\nabla T) - \sum_{i = \pm}\nabla\cdot(P_i\bu_i) - \sum_{i = \pm}q_i\rho_i\bu_i \cdot \nabla\phi.
\end{equation}
Here the first term in the right describes heat diffusion, the second term represents the work of thermal pressure converted into heat, and the third term that includes irreversible Joule heating describes the heat converted from electrostatic interactions. The derivation details for~\reff{th1eq} can be found in \emph{Supporting Information}. 

To describe thermally insulated electrodes, the boundary conditions of the temperature equation~\reff{th1eq} are prescribed by
\begin{align}\label{TBC}
\frac{\partial T }{\partial x} (\pm L) = 0.
\end{align}

\subsection{Governing equations}
In summary, we have the following governing equations to describe the thermal electrokinetics of EDLCs:
\begin{equation}
	\label{PNPFS}
	\left\{
	\begin{aligned}
		& -\nabla \cdot \ve_0\ve_r\nabla \phi = \sum_{i=\pm}q_{i}\rho_{i},  \\
		& \frac{\partial \rho_{\pm}}{\partial t} + \nabla\cdot\left( \rho_{\pm}\bu_{\pm}\right) = 0, \\ 
		& \rho_{\pm}\bu_{\pm} = -\frac{1}{\nu_{\pm}}\left( k_B T\nabla \rho_{\pm} +\frac{\partial P_{\pm}}{\partial T}\nabla T+ k_B T\frac{v\rho_{\pm}}{1-\sum_{i = \pm}v\rho_i}\sum_{j = \pm}\nabla \rho_j+q_{\pm}\rho_{\pm}\nabla \phi\right),\\
		& \frac{k_BC_0}{v}\frac{\partial T}{\partial t}  = \nabla\cdot (k\nabla T) - \sum_{i = \pm}\nabla\cdot(P_i\bu_i) - \sum_{i = \pm}q_i\rho_i\bu_i \cdot \nabla\phi, \\
		& P_\pm = k_BT\left(\rho_{\mp}-\frac{1}{v}\right)\log\left(1-\sum\limits_{j = \pm}v\rho_j\right).
	\end{aligned}
	\right.
\end{equation}
To nondimensionalize the equations, we introduce reference temperature $T_0$, microscopic length scale $\lambda_D = \sqrt{\frac{\ve_0\ve_r k_BT_0}{2e^2\rho_{\infty}}}$,  density $\rho_{\infty}$,  viscosity $\nu_0$, and time scale $\tau = \frac{\lambda_DL\nu_0}{k_BT_0}$. Also, we introduce dimensionless quantities $\widetilde{x}=x/L$, $\widetilde{t}=t/\tau$, $\widetilde{T}=T/T_0$, $\widetilde{v}=2v\rho_{\infty}$, $\widetilde{\phi}=\frac{e\phi}{k_{B}T_0}$, $\widetilde{k} = \frac{\tau k}{2k_B \rho_{\infty} L^2}$, and $\epsilon = \lambda_D/L$, which is often a small parameter for macroscopic EDLCs.  For binary monovalent electrolyte solutions under consideration, we assume $\nu_{\pm} = \nu_0$, consider salt density $\widetilde{c} = \frac{\rho_++\rho_-}{2\rho_{\infty}}$ and charge density $\widetilde{\rho} = \frac{\rho_+-\rho_-}{2\rho_{\infty}}$, and introduce $s = \log(1-\widetilde{v}\widetilde{c})$. Dropping the tildes in new variables, we have
\begin{equation}
\label{PNPFS}
\left\{
\begin{aligned}
&\frac{\partial c}{\partial t} = -\epsilon\frac{\partial j_c}{\partial x}, \quad j_c = -\left( \frac{T}{1-vc}  \frac{\partial c}{\partial x}+\rho \frac{\partial\phi}{\partial x}-\frac{2-vc}{v}s \frac{\partial T}{\partial x} \right), \\
&\frac{\partial \rho}{\partial t} = -\epsilon\frac{\partial j_{\rho}}{\partial x}, \quad j_\rho = -\left(T\frac{\partial \rho}{\partial x} +c \frac{\partial\phi}{\partial x} + T\frac{v\rho}{1-vc}\frac{\partial c}{\partial x}-\rho s\frac{\partial T}{\partial x}\right),\\
\end{aligned}
\right.
\end{equation}
where $j_{\rho}$ is the dimensionless Faradaic current scaled by $j_0 = 2e\rho_{\infty}\nu/Lk_BT_0$, and $j_c$ is the dimensionless salt current.
Thus, we have the following relations between the reduced variables and original density and velocity of each ionic species:
\begin{align}
	\rho_{\pm} = c\pm \rho,~~~u_\pm = \frac{j_c\pm j_{\rho}}{c\pm \rho}.
\end{align}
Finally, we have the following reformulated dimensionless governing equations for the electric potential, densities, and temperature:
\begin{equation}
\label{PNPFS-dimensionless}
\left\{
\begin{aligned}
& -\epsilon^2\frac{\partial^2\phi}{\partial x^2}= \rho, \\
&\frac{\partial c}{\partial t} = -\epsilon\frac{\partial j_c}{\partial x}, \quad j_c = -\left( \frac{T}{1-vc}  \frac{\partial c}{\partial x}+\rho \frac{\partial\phi}{\partial x}-\frac{2-vc}{v}s \frac{\partial T}{\partial x} \right), \\
&\frac{\partial \rho}{\partial t} = -\epsilon\frac{\partial j_{\rho}}{\partial x}, \quad j_\rho = -\left(T\frac{\partial \rho}{\partial x} +c \frac{\partial\phi}{\partial x} + T\frac{v\rho}{1-vc}\frac{\partial c}{\partial x}-\rho s\frac{\partial T}{\partial x}\right),\\
&\frac{C_0}{v} \frac{\partial T}{\partial t}  = k \frac{\partial^2 T}{\partial x^2} - \epsilon\sum_{i = \pm} \frac{\partial}{\partial x} (P_i u_i) - \epsilon j_{\rho} \frac{\partial \phi}{\partial x},\\
\end{aligned}
\right.
\end{equation}
where $s = \log(1-vc)$ and $P_{\pm}=\frac{v\rho_{\mp}-1}{v}T \log(1-\sum_{j = \pm}v\rho_j)$.

\subsection{Numerical Method}
\label{s:NumAlg}
The system~\reff{PNPFS-dimensionless} with boundary conditions~\reff{JBC}, \reff{RBC}, and \reff{TBC} is a numerically challenging problem, especially when boundary layers present due to small $\epsilon$ for macroscopic EDLCs.  For temporal integration, we propose a fully implicit time-discretization scheme
\begin{equation}
\label{PNPFS-timediscretization}
\left\{
\begin{split}
& -\epsilon^2 \frac{\partial^2 \phi^{n+1}}{\partial x^2} = \rho^{n+1}, \\
&\frac{c^{n+1}-c^{n}}{\Delta t} = -\epsilon \frac{\partial j_c^{n+1}}{\partial x}, \\
& j_c^{n+1} =-\left( \frac{T^{n+1}}{1-vc^{n+1}}  \frac{\partial c^{n+1}}{\partial x}+\rho^{n+1} \frac{\partial\phi^{n+1}}{\partial x}-\frac{2-vc^{n+1}}{v}s^{n+1} \frac{\partial T^{n+1}}{\partial x} \right), \\
&\frac{\rho^{n+1}-\rho^{n}}{\Delta t} = -\epsilon\frac{\partial j_{\rho}^{n+1}}{\partial x}, \\
& j_\rho^{n+1} = -\left(T^{n+1}\frac{\partial \rho^{n+1}}{\partial x} +c^{n+1} \frac{\partial\phi^{n+1}}{\partial x} + T^{n+1}\frac{v\rho^{n+1}}{1-vc^{n+1}}\frac{\partial c^{n+1}}{\partial x}-\rho^{n+1} s^{n+1}\frac{\partial T^{n+1}}{\partial x}\right),\\
&\frac{C_0}{v} \frac{T^{n+1}-T^{n}}{\Delta t}  =  k\frac{\partial^2 T^{n+1}}{\partial x^2} - \epsilon\frac{\partial}{\partial x}\left(  P_+^{n+1} u_+^{n+1}+P_-^{n+1} u_-^{n+1}\right) -\epsilon j_{\rho}^{n+1}\frac{\partial \phi^{n+1}}{\partial x},
\end{split}
\right.
\end{equation}
where $\Delta t$ is a time step size, and $c^{n}$, $\rho^{n}$, $T^{n}$, and $\phi^{n}$ are numerical approximations of $c(\cdot, t_n)$, $\rho(\cdot, t_n)$, $T(\cdot, t_n)$, and $\phi(\cdot, t_n)$ at $t_n = n\Delta t$, respectively.

After temporal discretization, the system~\reff{PNPFS-timediscretization} with boundary conditions is spatially discretized with a collocation method implemented in BVP4C~\cite{BVP4C} in the Matlab. For small $\epsilon$, boundary layers are resolved with an adaptive, nonuniform mesh, in which grid points are  densely distributed adjacent to the electrodes. The resulting nonlinear algebraic equations are solved iteratively by Newton-type methods. A strategy of continuation on the parameter $\epsilon$ is adopted to generate good initial guesses for iterations at the first time step.  Further numerical details are presented in the \emph{Supporting Information}.

\section{Results}
\label{s:Results}
To demonstrate the performance of the proposed model, extensive numerical simulations based on the model are conducted to investigate the effect of charging-discharging scan rates and ionic sizes on the thermal electrokinetics of EDLCs and unravel underlying mechanism of the reversible and irreversible heat generation. In the periodic charging-discharging processes, a time-dependent voltage is applied through the boundary condition~\reff{RBC} with the scheme~\reff{CVpsi}, in which the maximum and minimum surface electric potentials are set as $\psi_{max} = 1 V$ and $\psi_{min} = 0 V$. We consider monovalent binary electrolyte solutions that are blocked by two thermally insulated electrodes.  Unless specified otherwise, the electrolyte relative permittivity is taken as that of water $\epsilon_r = 66.1$, uniform ionic diffusion coefficients $D_{\pm} = k_BT_0/ \nu_{\pm} = 1.7\times 10^{-10}m^2/s$ are considered, and typical solvated ion diameter $a$ is used, ranging from $0.50$ to $0.68$ nm~\cite{pilon_jps14}. The initial and bulk ion concentrations are set as $\rho_{\infty} = 1 M$, which gives $\lambda_D=0.3 nm$. To simulate macroscopic EDLCs, we consider a computational domain $[-L, L]$ with $L=1\mu m$. Therefore, $\epsilon=0.0003$ which indicates that there are thin boundary layers closed to electrodes.

\subsection{Charging dynamics}

\begin{figure}[htbp]
	\label{f:C+Psi+I}
	\centering
	\includegraphics[scale=.7]{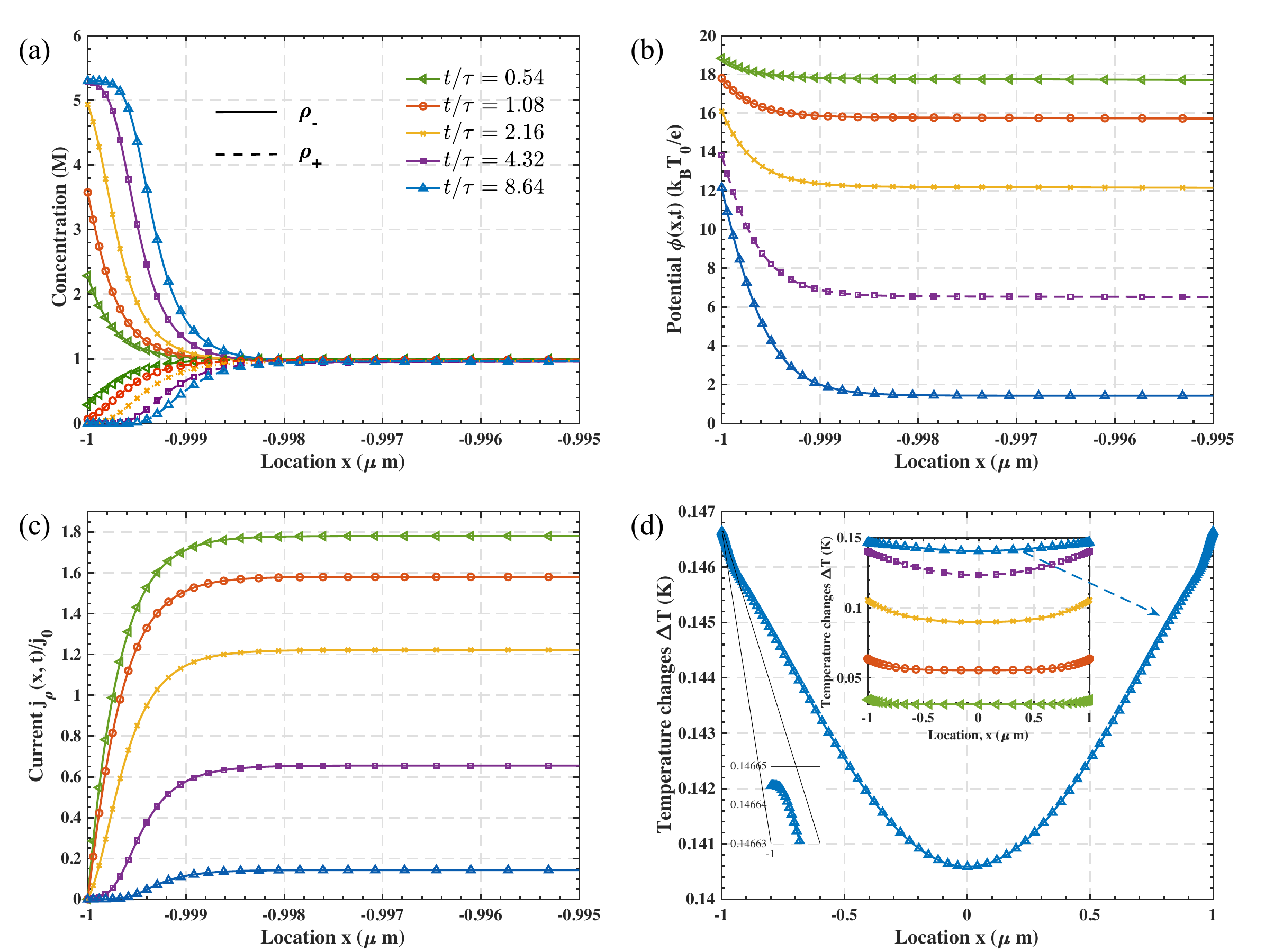}
	\caption{  Snapshots of ionic concentration $\rho_{\pm}(\bx,t)$ profiles (a), electric potential $\phi(\bx,t)$ profiles (b), current density $j_{\rho} (\bx,t)$ profiles (c), and temperature $T(\bx,t)$ profiles (d).   }
\end{figure}

The proposed model is first applied to study thermal electrokinetics of an EDLC charging to steady states. A fixed surface potential $\phi_s(t) \equiv 20  \left( k_BT_0/e \right)$ is applied at electrodes; cf. the Robin boundary conditions~\reff{RBC}. The simulations take a uniform ionic size $a = 0.68 nm$, which implies a saturation concentration  $\rho_{max} = 1/a^3 = 5.3 M$.  As displayed in Fig.~\ref{f:C+Psi+I} (a),  counterions gradually accumulate next to the surface with the applied voltage, and the concentration eventually reaches a saturation concentration due to steric hindrance. By contrast, coions are depleted away from the electrode due to the electrostatic interaction. As the charging process reaches a steady state,  a pronounced saturated layer with uniform concentration $\rho_{-} = \rho_{max}$ forms near electrodes.  Fig.~\ref{f:C+Psi+I} (b) depicts the profiles of electrostatic potential as charging process proceeds. Notice that the potential first drops linearly from the surface value $\phi_s$ within the Stern layer $H$. In addition, the electrostatic potential further gets screened quickly because of the attracted counterions in the EDL.  

Fig.~\ref{f:C+Psi+I} (c) shows the current density distribution $j_{\rho}(\bx,t)$ (rescaled by $j_0$) during the charging process. Since zero-flux boundary conditions are imposed, the current density vanishes in the Stern layer for the entire charging process. In initial stages, there is large current density extending into bulk regions, ascribing to the transport of the attracted of counterions and repulsed of coions.  As the saturation layer builds up, the electrostatic interactions get screened and the  current both in EDL and bulk decreases. Eventually, the current vanishes everywhere when the system approaches the steady state. Overall, it should be emphasized that ionic concentrations, potential, and current all level off quickly to bulk values within a very thin boundary layer next to the electrode (about $0.002 \mu $m width). The inset of Fig.~\ref{f:C+Psi+I} (d) shows that the profiles of temperature $T(\bx,t)$ exhibit a parabolic shape for the whole domain during the charging process. The temperature at boundaries takes the lead to rise and the internal bulk electrolytes subsequently get heated up as well. With thermally insulated boundary conditions,  the temperature becomes homogeneous everywhere in the steady state. 
\subsection{Periodic charging and discharging}

\begin{figure}[htbp]
	\label{f:dyn_CIT}
	\centering
	\includegraphics[scale=.8]{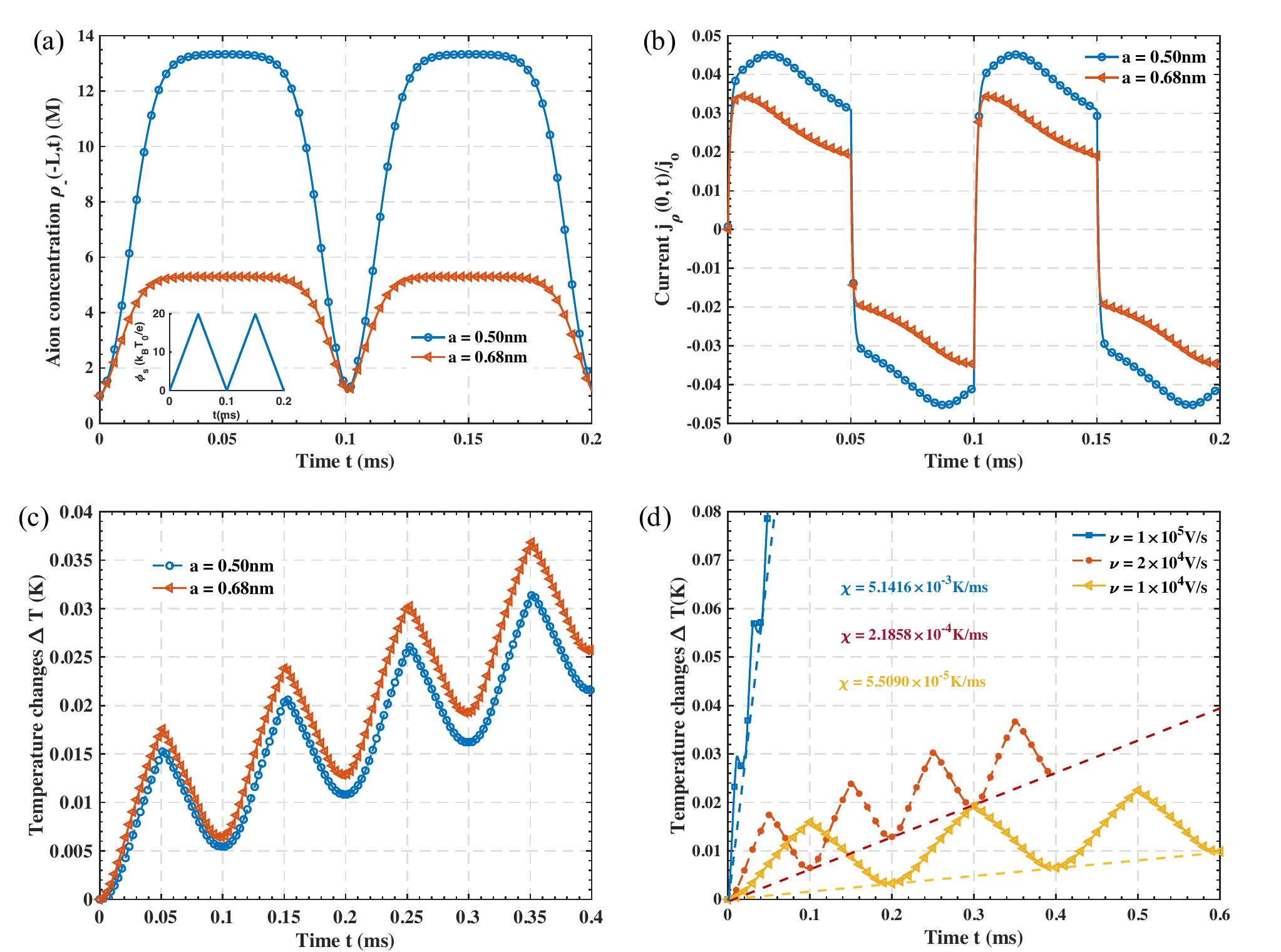}
	\caption{   Evolution of counterion concentration $\rho_{-}(-L,t)$ (a), current density $j_{\rho}(0,t)$ in the bulk (b),  and temperature change $\Delta T = T(-L, t)- T(-L, 0)$ (c), using the cyclic voltammetry scan rate $\nu = 2\times10^4V/s$ and different ionic diameters $a=0.50$ nm and $a=0.68$ nm.  (d) Temperature change $\Delta T$ at electrodes using different scan rate $\nu$ and an ionic diameter $a = 0.68$ nm. The parameter $\chi$ denotes the slope of dash lines that reflect the linear increase of $\Delta T$ with respect to time. }
\end{figure}


The proposed model is also applied to probe thermal electrokinetics in EDLCs with a linearly varying surface potential $\phi_s$ given by~\reff{CVpsi}, in which $\phi_{min}=0 (k_BT_0/e)$ and $\phi_{max} = 20 (k_BT_0/e)$; cf. the inset plot of Fig.~\ref{f:dyn_CIT} (a). Various ionic sizes and cyclic voltammetry scan rates are considered to evaluate their impact on the thermal electrokinetics.  As shown in Fig.~\ref{f:dyn_CIT} (a), the counterion concentration at the Stern/diffuse layer interface increases synchronously as the potential rises, and reaches their maximum allowable saturation values that are determined by their ionic sizes. As the potential decreases, the concentration declines synchronously back to the initial bulk values at the end of cycles. Fig.~\ref{f:dyn_CIT} (b) presents the periodic evolution of current density in the bulk region. Obviously, one can observe that the current surges drastically to a maximum value, and then decreases gradually, in that electrostatic attractions, i.e., the driving force, are weakened by the accumulated ions. Comparing two current curves, one can see that the smaller ionic size has larger maximum flux, since more ions are allowed to transport into EDLs. After $t=0.05$ ms, the potential begins to decrease linearly and the attracted ions in the EDL start to diffuse back to the bulk.  This explains the abrupt switching of the positive charge current to negative charge current at the same time. In addition, the shape of the profile before and after $t=0.05$ ms is highly symmetric. 

In addition, simulations are performed to assess the impact of ionic size on the temperature in continous charging-discharging cycles.  From Fig.~\ref{f:dyn_CIT} (c), one can see that the temperature at the Stern/diffuse layer interface rises in a sawtooth shape. It is heated up monotonically in a charging stage, and partially cooled down in a discharging stage. This demonstrates that the proposed model is able to capture the reversible and irreversible heat generation in the charging-discharging cycles. Overall, the temperature still rises oscillatory due to the irreversible contribution. Furthermore, it is seen that a larger ionic size contributes more irreversible heat, leading to overall steeper temperature rise. To further quantify the temperature rise, we use $\chi$ to denote the slope of overall temperature rise; cf. Fig.~\ref{f:dyn_CIT} (d). This figure presents the effect of scan rate of $\phi_s$ on the temperature rise. It is found that, when the scan rate changes from $\nu = 1\times10^4 V/s$ to $\nu = 2\times10^4 V/s$, the slope $\chi$ is correspondingly magnified by four times. Further increase from $\nu = 2\times10^4 V/s$ to $\nu = 1\times10^5 V/s$ reveals the scaling relation $\chi \propto \nu^2$. To understand this, we define the average current  $\bar{I}$ of a voltammetry cycle  as  
\begin{align}
	\bar{I} = \frac{\int_{t = 0}^{\tau_{cv}}j_{\rho}(0, t) dt}{\tau_{cv}},
\end{align}
where $\tau_{cv}$ is the half cycle period. Numerical results find that the average current $\bar{I}$ scales linearly with the increase of $\nu$. Therefore, it comes to the conclusion that the contribution of irreversible heat, represented by $\chi$, is proportional to the square of the current, i.e., $\chi \propto \bar{I}^2$. Such a conclusion is consistent with previous understanding on Joule heating effect~\cite{Pilon_JPS15, Zhang_TA16, Janssen_PRL17}.

\subsection{Heat generation}

\begin{figure}[htbp]
	\label{f:compare}
	\centering
	\includegraphics[scale=.68]{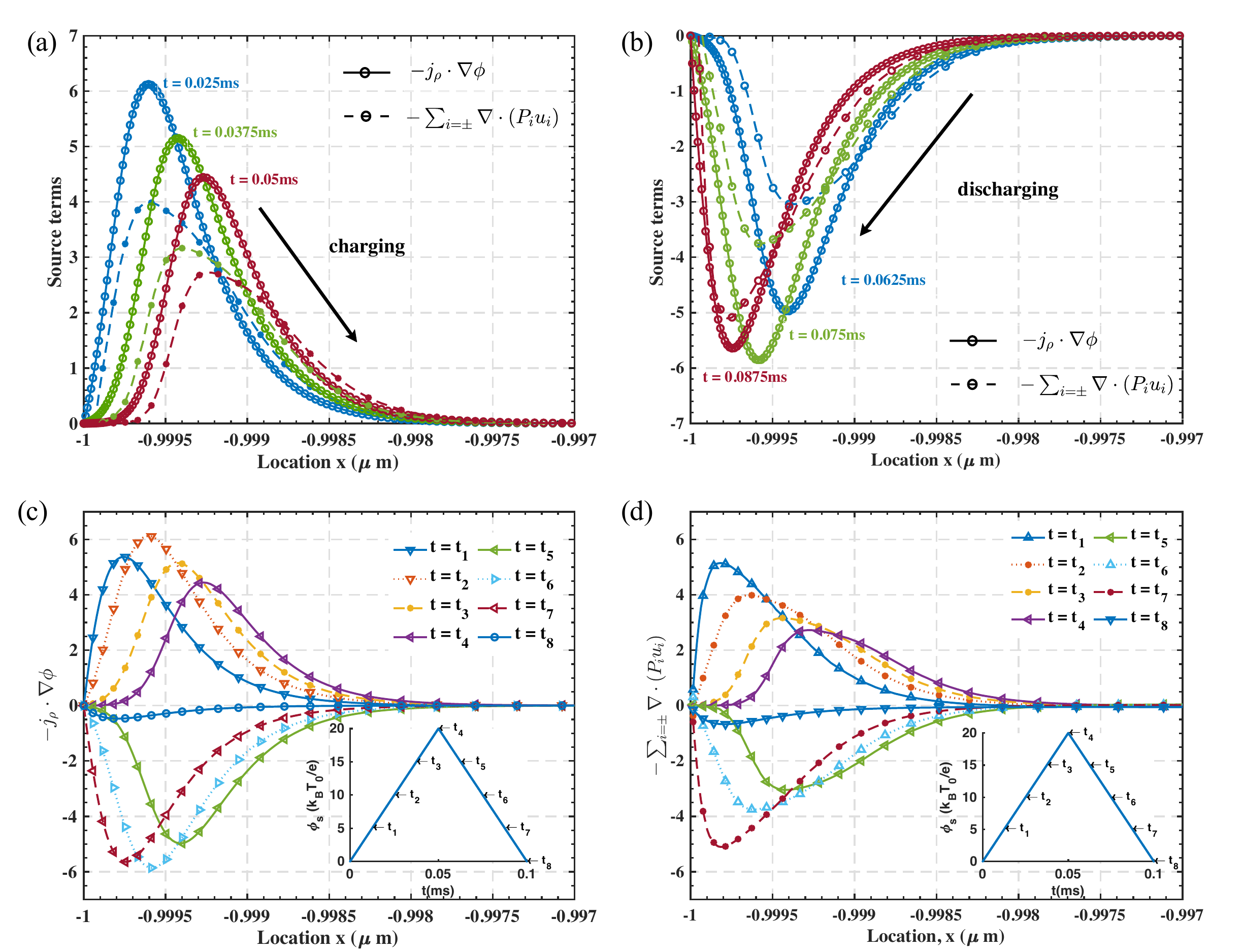}
	\caption{  Comparison of two heat generation source terms in the charging (a) and discharging (b) stage.   Evolution of the source terms $- j_{\rho}\cdot\nabla\phi$ (c) and $-\sum_{i = \pm}\nabla\cdot(P_iu_i)$ (d) in one charging-discharging cycle. }
\end{figure}

To further understand the heat generation of the proposed model, we investigate contributions of two source terms of the temperature evolution equation in~\ref{PNPFS-dimensionless}: $-j_{\rho}\cdot\nabla\phi$ due to the charge current and $-\sum_{i = \pm}\nabla\cdot(P_iu_i)$ due to the thermal pressure. Fig.~\ref{f:compare} (a) and (b) compare such two terms in the charging and discharging stage.  First, both terms are positive in the charging stage and negative in the discharging stage, indicating that they are both exothermic and endothermic in charging and discharging stages, respectively. In addition, the contribution made by the term due to charge current is a bit more significant than the term due to thermal pressure in the whole charging-discharging cycle.  It is interesting to note that such terms both vanish at the boundary, and mainly concentrate and peak in the boundary layers.  Although such terms are small in the bulk region, the bulk region is much larger than the boundary layers and therefore they still contribute a fair amount of heat in charging-discharging cycles.  

In order to understand the dynamics of such terms, we also plot the distribution of such terms in various phases in one charging-discharging cycle. Fig.~\reff{f:compare} (c) and (d) display $8$ snapshots of the distribution of the terms at times that divide the cycle into $8$ phases; cf. the labels in the inset.  The source term due to charge current rises positively and reaches the maximum value at the first quarter of the cycle. After that, the profile lowers with the position of the peaks moving away from the electrode, and suddenly changes its sign in the second half of the cycle. For the second half, i.e., the discharging process, the profile follows a similar pattern, rising negatively in the third quarter and diminishing in the fourth quarter.  For the term due to thermal pressure, the profiles in Fig.~\ref{f:compare} (d) show a different pattern---it reaches the maximum in the first quarter, then damps in the rest of charging process, and keeps rising negatively in discharging process until the seven eighths of one cycle. Notice that the peaks of both terms move away from the electrode in the charging process and approach the electrode in the discharging process. This can be explained by the fact that the accumulated counterions are piling up in layers at electrodes due to steric hindrance in the charging process, and the layering counterions dissolve back to the bulk in the discharging process. 


\begin{figure}[htbp]
	\centering
	\includegraphics[scale=.68]{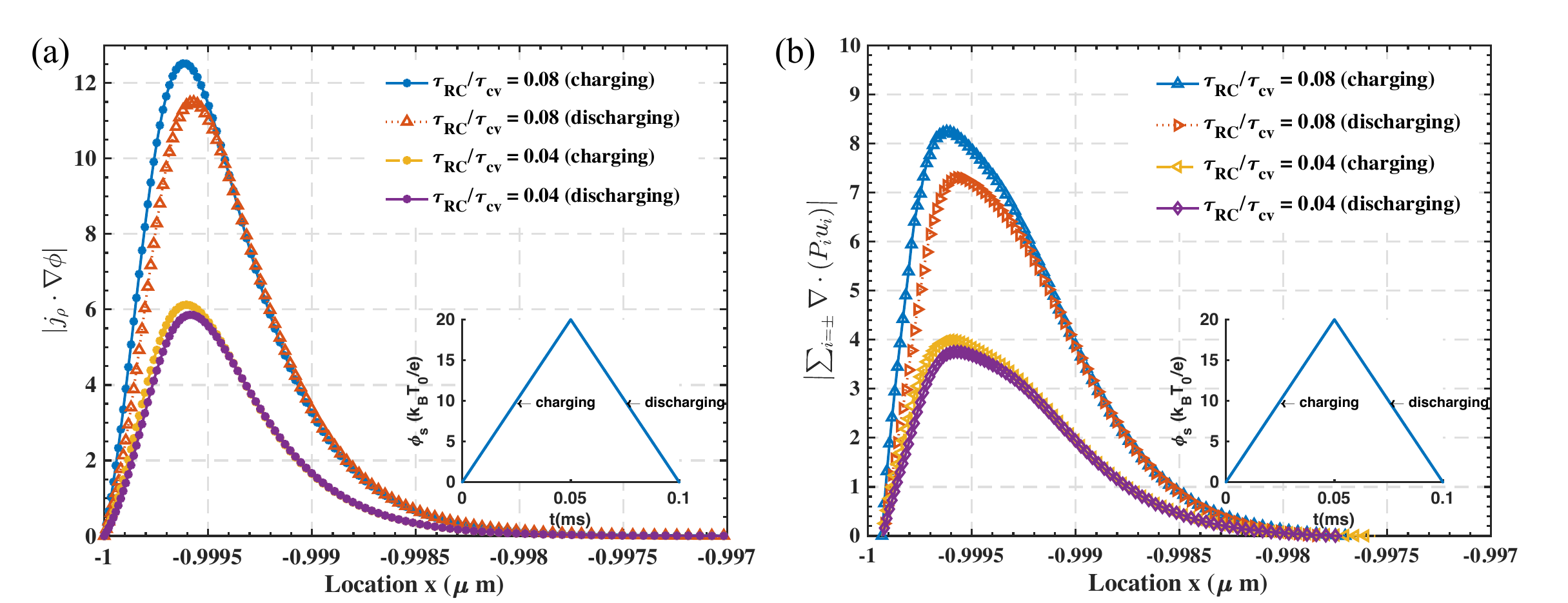}
	\caption{   The magnitude of heat generation source terms $j_{\rho}\cdot\nabla\phi$ (a) and $\sum_{i = \pm}\nabla\cdot(P_iu_i)$ (b) with different scan rates of surface potential in the middle of charging and discharging stages. 
	}\label{f:sources}
\end{figure}


To understand the impact of scan rate on the heat generation,  simulations are also performed with different scan rates that are represented by $\tau_{cv} = \frac{\psi_{max}-\psi_{min}}{\nu}$. It has been systematically studied that the dimensionless ratio $\tau_{RC}/\tau_{cv}$ is a key parameter to characterize the CV measurements~\cite{GirardWangPilon_JPCC15, LinLian_PRL22}. Here $\tau_{RC} = \frac{\lambda_DL\nu_0}{k_BT_0}$ is the characteristic relaxation timescale.  As shown in Fig.~\reff{f:sources} (a), the difference of the magnitude of $j_{\rho}\cdot\nabla\phi$ in the charging and discharging stages is magnified as the timescale ratio increases. Similar results can be observed for the heat source term $|\sum_{i = \pm}\nabla\cdot(P_iu_i)|$. The difference of the magnitude in charging and discharging processes is related to the irreversible heat generation in EDLCs. Such results are consistent with the results shown in Fig.\ref{f:dyn_CIT} (d) that temperature rise over cycles due to irreversible heat generation is enhanced significantly as the scan rate increases. 

\subsection{Cyclic voltammetry}

\begin{figure}[htbp]
	\centering
	\includegraphics[scale=.68]{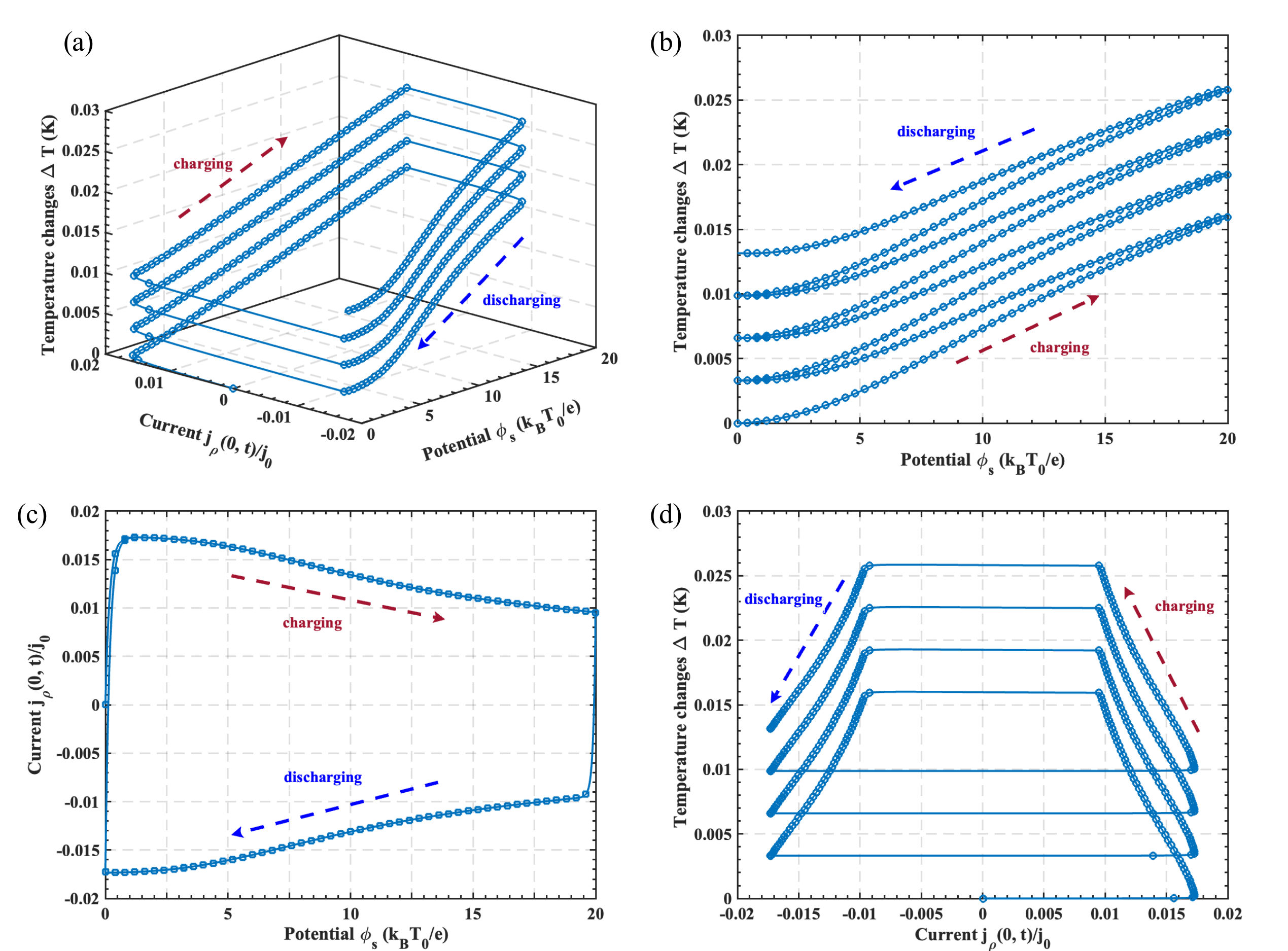}
	\caption{  (a) Evolution curve of the charge current in the bulk $j_{\rho}(0, t)$, the electrostatic surface potential $\phi_s(t)$, and the temperature change $\Delta T$ compared to the initial temperature at electrodes in cyclic voltammetry measurements; (b) Projection of the evolution curve to the $\phi_s$-$\Delta T$ plane; (c)  Projection of the evolution curve to the $\phi_s$-$j_{\rho}$ plane; (d) Projection of the evolution curve to the $j_{\rho}$-$\Delta T$ plane.}
	\label{f:cvt}
\end{figure}


Cyclic voltammetry is a very powerful method to characterize the performance of electrochemical devices under various conditions~\cite{WangPilon_EA12,WangPilon_JPCC13,GirardWangPilon_JPCC15,GirardPilon_EA16,MeiPilon_EA17}.
Here, we apply the proposed model to understand the interplay between temperature and other key factors in CV measurements.  Again, the boundary conditions~\reff{RBC} with the surface potential given by the scheme~\reff{CVpsi} are prescribed with the Poisson's equation. Fig.~\ref{f:cvt} (a) presents a 3D evolution curve of the charge current in the bulk $j_{\rho}(0, t)$, the electrostatic surface potential $\phi_s(t)$, and the temperature change $\Delta T$ compared to the initial temperature at electrodes, in several periods of charging and discharging. In contrast to the traditional plot of current-versus-potential, such a CVT presentation includes the temperature as another player and unravels the dependence of the temperature on both the charge current and surface potential. Overall, one can observe that the temperature rises in a spiral path. It is clearly seen that the temperature rises significantly and linearly after the charge current climbs over its peak value, and remains unchanged in the transitions between charging and discharging processes. 

To further understand the pairwise interplay between two of three factors, the CVT evolution curve is projected to the $\phi_s$-$\Delta T$ plane, $\phi_s$-$j_{\rho}$ plane, and $j_{\rho}$-$\Delta T$ plane in Fig.~\ref{f:cvt}. The dependence of $\Delta T$ on $\phi_s$ shows clear alternation between increasing and switching back, with a net increase about $0.003$ K in each cycle. The magnitude is smaller than reported experimental data for commercial EDLCs~\cite{Schiffer_JPS06, pilon_jps14, Zhang_TA16, Parvini_IEEE16}. There are several possible reasons for the discrepancy, e.g., simple planar geometry rather than porous structures is employed in the model.  The projection onto the $\phi_s$-$j_{\rho}$ plane gives a typical CV curve that is often shown in CV simulations. It should be pointed out that the projection of the 3D curve of several cycles almost collapse into one overlapped CV curve, indicating that the rising of temperature does not impact the CV simulations very much. One possible reason is that the diffusion coefficient, dielectric coefficient, etc. are assumed to be independent of temperature in the model. Further refinement of our model on this respect should be considered in future work.  In addition, the CV curve depicts that the current rises drastically in the initial short period of time. After reaching the peak value, the current declines gradually until it transits from charging process to the discharging process.  The peak value can be explained by the formation of a saturation layer in the EDL~\cite{WangPilon_EA12}. The dependence of $\Delta T$ on $j_{\rho}$ further confirms our finding that the temperature only changes significantly when the charge current is large and has no variation during the charging-discharging transitions. 


\begin{figure}[htbp]
	\centering
	\includegraphics[scale=.68]{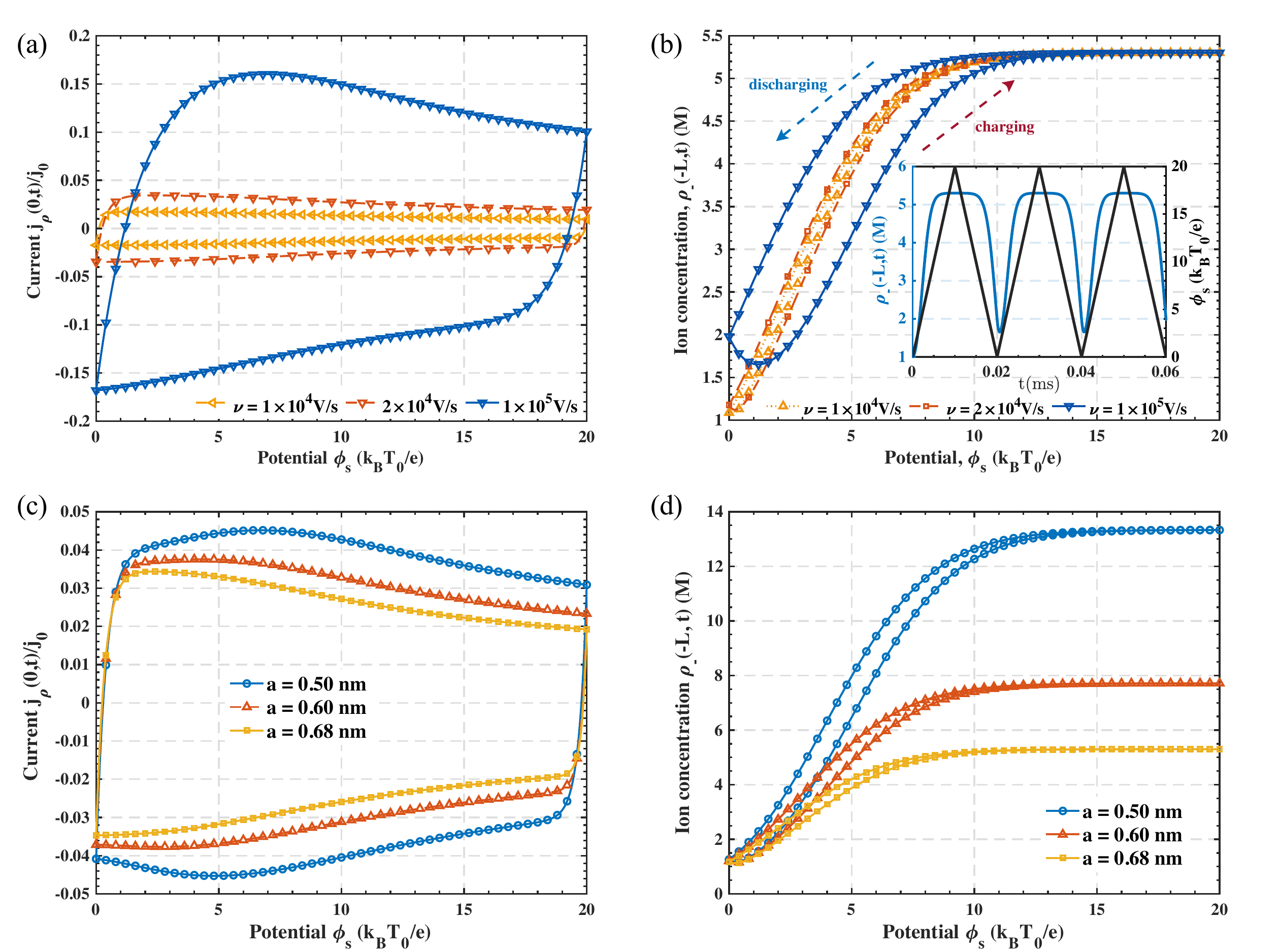}
	\caption{  The $j_{\rho}$-$\phi_s$ curve (a) and $\rho_{-}(-L, t)$-$\phi_s$ curve (b) in CV simulations with various scan rates in one cycle.  The inset plot gives the surface potential and counterion concentration against time. The $j_{\rho}$-$\phi_s$ curve (c) and $\rho_{-}(-L, t)$-$\phi_s$ curve (d) in CV simulations with various ionic sizes in one cycle.  
	}	\label{f:CV}
\end{figure}


In order to further probe the CV measurements, we perform simulations with different scan rates and ionic sizes.  Fig.~\ref{f:CV} (a) displays the CV diagram with $\psi_{max} = 40 k_BT_0/e$, $a = 0.68nm$, and scan rates ranging from $\nu = 1.0\times 10^4 V/s$ to $\nu = 1\times 10^5 V/s$.  With faster scan rates, the range for charge current in the charging-discharging process becomes much larger. Also, it is known that the enclosed area of the CV curve represents the charge per unit surface area accumulated at the electrode surface during one cycle~\cite{WangPilon_EA12}. Thus, faster scan rates gives larger integral capacitance for EDLCs in CV measurements. Fig.~\ref{f:CV} (b) presents the dependence of the counterion concentration at electrodes on the surface potential. Of interest is that, as the scan rate increases, the charging and discharging processes follow different pathways, exhibiting a \emph{hysteresis} diagram.  When the system relaxation time scale is much larger than that of scanning potential, i.e. $\tau_{RC} \gg \tau_{cv}$,  the thermal electrokinetics in EDLCs cannot follow the change of scanning potential and the concentration show obvious delayed dynamics. Starting from the second cycle, the discharging stage ends before the concentration relaxes to the bulk concentration $1$ M. The retard concentration keeps declining a little bit even though the potential is increasing in the beginning of the next charging stage. This is confirmed by the inset plot of the surface potential and counterion concentration against time. It is reasonable to see that the delayed dynamics becomes less obvious for a slower scan rate $\nu = 1.0\times 10^4 V/s$, for which $\tau_{RC}/\tau_{cv} \sim \mathcal{O} (1)$. The corresponding concentration relaxes closer to the bulk concentration at the charging-discharging transition. 

Fig.~\ref{f:CV} (c) and (d) present the CV simulations with a fast scan rate $\nu = 1\times 10^5 V/s$ and three different ionic sizes ranging from $a = 0.6$ nm to $a = 0.68$ nm.  With a small ionic size, the range of charge current enlarges, being consistent with the understanding that the maximum value (hump) in the curve is due to the formation of a saturation layer in the EDL~\cite{WangPilon_EA12}. Therefore, a larger area is enclosed by the CV curve and higher integral capacitance is achieved in the CV simulations with a smaller ionic size. The counterion concentration at the electrode shown in Fig.~\ref{f:CV} (d) demonstrates that, with a smaller ionic size, the saturation concentration increases and the area enclosed by the hysteresis-loop curve gets larger as well. This can be ascribed to the fact that more counterions are involved in the periodic formation and dissolution of the EDLs. 
\section{Conclusions}
\label{s:conclusions}
This study has proposed a new variational, thermodynamically consistent model to predict thermal electrokinetics in EDLCs by using an energetic variational approach. Modified Nernst-Planck (NP) equations incorporating the diffusion, ionic steric effect, and convection due to the gradient of temperature and electrostatic potential have been developed by using the least action principle and maximum dissipation principle. The proposed modified NP equations are a thermodynamically consistent generalization of the steric NP equations presented in~\cite{Bazant_PRE07II} to non-isothermal scenarios with temperature inhomogeneity. Temperature evolution equations with heat source due to thermal pressure and electrostatic interactions have been derived by using laws of thermodynamics. 

Extensive simulations of EDLCs have demonstrated that the developed model can successfully predict temperature oscillation in the charging-discharging processes, and larger ionic sizes and faster scan rate of surface potential lead to faster oscillatory temperature rise. In addition, the temperature rise slope, reflecting the irreversible Joule heating effect, has been found to scale quadratically with the average current in cyclic voltammetry (CV). Further study on heat generation source terms have unraveled that they are both exothermic and endothermic in charging and discharging stages, respectively. The discrepancy of source terms in charging and discharging stages, due to irreversible heat generation, gets larger as the scan rate increases.  In CV measurements, the temperature has been included as an additional dimension to the traditional CV curve, resulting in a 3D spiral CVT curve. For fast scan rates, the CV curves and counterion concentrations have evidenced that the thermal electrokinetics in EDLCs cannot follow the scanning potential in time, showing delayed dynamics with hysteresis diagrams. Such simulation results have demonstrated that the proposed model is capable of predicting thermal electrokinetics in EDLCs during charging/discharging processes. With further refinement on the description of porous structure of electrodes and temperature dependent parameters, the proposed model is expected to provide a useful tool for the design, optimization, and manufacture of EDLCs with safety and long-term stability.



\vspace{4 mm}

\noindent{\bf Acknowledgments.}
C. Liu's work is partially supported by NSF grants DMS-1950868 and DMS-2118181. P. Liu's work is partially supported by NSF grants DMS-1816740. S. Zhou's work is partially supported by the National Natural Science Foundation of China 12171319, Natural Science Foundation of Jiangsu Province (BK20200098), China, Young Elite Scientist Sponsorship Program by Jiangsu Association for Science and Technology, and Shanghai Science and Technology Commission (21JC1403700)

}
\bibliography {Latest}
\bibliographystyle{plain}

\newpage
\begin{center}
\textbf{\Large {\it Supplemental Materials:} }
\bigskip
\\\textbf{ \Large Energetic Variational Approach for Prediction of Thermal Electrokinetics in Charging and Discharging Processes of Electrical Double Layer Capacitors}
\end{center}
\setcounter{equation}{0}
\setcounter{figure}{0}
\setcounter{table}{0}
\setcounter{page}{1}
\setcounter{section}{1}
\makeatletter
\renewcommand{\theequation}{S\arabic{equation}}
\renewcommand{\thefigure}{S\arabic{figure}}

\subsection{Derivation of conservative forces}\label{sec:level1} 
		
		In this section, we present mathematical details on the derivation of \reff{con} in the main text. According to the Least Action Principle (LAP)~\cite{BobHyonLiu_JCP10, Deriglazov, Biot},  conservative forces can be obtained by taking the first variation of an action functional  $A(\bx(\bX,t))$ with respect to the flow map $\bx(t) = \bx(\bX ,t)$; cf. its definition~\reff{FlowMap} in the main text. 
		Introduce a perturbation $\bx(t)+\ve \by(t),$ where $\ve$ is an infinitesimal number and $\by(t)$ is an arbitrary smooth function with compact support. 
		Then the conservative force $\bm{F}_{con}$ in the Lagrangian coordinate is defined through
		\[
		\begin{aligned}
			\delta A\left(\bx \right) &=\frac{d}{d \ve}\bigg| _{\ve = 0} A\left( \bx + \ve\by\right)  = \int_{t = 0}^{t^*}\int_{\Omega_0^{X}}\left[ \bm{F}_{con} \right]\cdot \by \,d\bX dt,	 	
		\end{aligned}
		\]
		where $t^*>0$ denotes the elapsed time and $\Omega_0^{X}$ is the Lagrangian reference domain of $\Omega_t^{x}$. The conservative force in the Eulerian coordinate can be defined analogously.
		
		For binary electrolytes under consideration, there are two flow maps $\bx_{\pm}(\bX ,t)$ corresponding to the motion of cations and anions, respectively.
		Then,  the corresponding action functional is defined as 
		\[
		\begin{aligned}
			A\left(\bx_{+}(\bX, t),  \bx_{-}(\bX, t)\right) &= -\int_{0}^{t^*}  \left( F_{pot}(\Omega, t) + F_{ent}(\Omega, t) \right) dt \\
			&:= A_{pot}\left( \bx_+(\bX,t),  \bx_{-}(\bX, t)\right) +A_{ent}\left( \bx_+(\bX,t),  \bx_{-}(\bX, t)\right).
		\end{aligned}
		\]
		Here, the mean-field electrostatic free-energy functional $F(V, t)$ is split into $F_{pot}(V, t)$ and $F_{ent}(V, t)$.
		As shown in~\reff{ElePot}, the electrostatic potential energy is given by
		\[
		\begin{aligned}
			F_{pot}(V,t) &= \sum_{i,m = \pm}\frac{q_iq_m}{2}\iint_{V}\rho_i(\bx,t)\rho_m(\bx^{\prime},t)G(\bx,\bx^{\prime})d\bx d\bx^{\prime} \\   
			&\qquad + \sum_{i= \pm} q_i\int_{V}\rho_i(\bx)\left( \psi_X(\bx,t) + \int_{\Omega \backslash V} \sum_{m = \pm} \rho_m(\bx^{\prime}, t)G(\bx,\bx^{\prime})d\bx^{\prime}\right) d\bx.
		\end{aligned}
		\]
		As shown in~\reff{ent_re}, the entropy contribution is given by
		\[
		\begin{aligned}
			F_{ent}(V,t) &= \int_{V} \Psi\left( \rho_+(\bx,t),\rho_-(\bx,t), T(\bx,t) \right)  d\bx  \\
			&= \int_{V} \left[ \sum_{i = \pm}\Psi_i\left(\rho_{i}(\bx, t), T\left(\bx, t \right)  \right)  + \Psi_0\left( \rho_{+}(\bx,t), \rho_{-}(\bx, t), T(\bx, t)\right) -\Psi_T(T(\bx, t))\right] d\bx,
		\end{aligned}
		\]
		where $\Psi_{\pm}, \Psi_0$, and $\Psi_T$ are defined as
		\[
		\begin{aligned}
			\label{ent_part}
			\Psi_{\pm}\left(\rho_{\pm}(\bx, t), T\left(\bx, t \right)  \right)  &:= k_BT(\bx,t) \rho_{\pm}(\bx,t)\log\left( v\rho_{\pm}(\bx,t)\right) ,\\
			\Psi_0\left(\rho_{+}(\bx, t), \rho_{-}(\bx, t), T\left(\bx, t \right)  \right) &:= k_BT(\bx,t)\frac{1-\sum\limits_{i = \pm}v\rho_i(\bx,t)}{v}\left[\log(1-\sum_{i = \pm}v\rho_i(\bx,t))\right] ,\\
			\Psi_T\left( T(\bx, t)\right)  &:= \frac{k_BC_0}{v}T(\bx,t)\log T(\bx,t).
		\end{aligned}
		\]
		The action $A_{ent}$ from the entropy, as a functional of two flow-maps, is given by,
		\begin{eqnarray}\label{Aent}
			&&A_{ent}[\bx_+(\bX,t), \bx_-(\bX,t)] \nonumber\\ &=& - \int_0^{t^*} \int_{\Omega_t^x} \left[ \Psi_+\left(\rho_+(\bx,t),T(\bx,t) \right) + \Psi_-(\rho_-(\bx,t), T(\bx,t) ) \right. \nonumber \\
			&& \qquad \qquad \qquad +\left. \Psi_0(\rho_+(\bx,t) ,\rho_-(\bx,t) ,T(\bx,t) ) - \Psi_T(T(\bx,t) ) \right] d\bx dt \nonumber \\
			&=& - \int_0^{t^*} \int_{\Omega_0^X} \Bigg[ \Psi_+\left(\frac{\rho_+(\bX)}{J_+} ,T_+(\bX)\right) +\Psi_0\left(\frac{\rho_+(\bX)}{J_+ }, \rho_-(\bx_+(\bX,t),t), T_+(\bX)\right) \nonumber\\
			&&   \qquad \quad+ \Psi_T(T_+(\bX)) \Bigg] J_+ d\bX dt  - \int_0^{t^*} \int_{\Omega_0^X}  \Psi_-\left(\frac{\rho_-(\bX)}{J_- },T_-(\bX) \right) J_- d\bX dt.
		\end{eqnarray}
		Here the densities depend on the flow-maps through the deformation gradient tensor $\mathcal{F}_{\pm} (\bX,t) =  \frac{\partial \bx_{\pm}(\bX,t)}{\partial \bX}$ with the Jacobians defined by $J_{\pm} = \det \mathcal{F}_{\pm}$. We focus on the mechanic part of the system and assume that $T(\bx,t)$ follows both the flow maps through $T(\bx_+(\bX,t),t) =T_+(\bX)$ and $T(\bx_-(\bX,t),t) = T_-(\bX)$. The Eq.~\reff{Aent} is used to calculate the entropic part of the conservative forces for $\rho_+$. 

		Now we take the derivation of conservative forces for cations as an example. Let  $\by(\bX,t) = \tilde{\by}(\bx_+(\bX,t),t )$ be a perturbation of the flow map $\bx_+(\bX,t)$.
		The variation of the action $A_{ent}$ with respect to the flow map $\bx_{+}(\bX ,t)$ leads to
		\begin{eqnarray}
			&& \left. \frac{d}{d \epsilon} \right|_{\epsilon =0} A_{ent}[\bx_+(\bX,t)+\epsilon \by(\bX,t) , \bx_-(\bX,t) ] \nonumber \\ &=&  - \int_0^{t^*} \int_{\Omega_0^X} \left[ \left(\Psi_+ + \Psi_0 -\frac{\partial (\Psi_+ +\Psi_0)}{\partial \rho_+}  \frac{\rho_+}{J_+}\right) Tr(\mathcal{F}_+^{-1} \frac{\partial \by}{\partial \bX} ) +\frac{\partial \Psi_0}{\partial \rho_-} \frac{\partial \rho_-}{\partial \bx } \cdot \by\right]J_+ 
			d\bX dt \ \nonumber\\ &=& \int_0^{t^*} \int_{\Omega_t^x} \left[ \nabla \left(\Psi_+ + \Psi_0 -\frac{\partial (\Psi_+ +\Psi_0)}{\partial \rho_+}  \frac{\rho_+}{J_+}\right) - \frac{\partial \Psi_0}{\partial \rho_-} \frac{\partial \rho_-}{\partial \bx } \right] \cdot \by d\bx dt.
		\end{eqnarray}
		Here we use the fact that
		\begin{equation}
			\begin{aligned}
			\left. \frac{d}{d \epsilon} \right|_{\epsilon =0} \left|\frac{\partial (\bx_+ + \epsilon \by)}{\partial \bX} \right| &=  \left|\frac{\partial \bx_+ }{\partial \bX} \right| \left. \frac{d}{d \epsilon} \right|_{\epsilon =0}  \left| I + \epsilon \left(\frac{\partial \bx_+}{\partial \bX}\right)^{-1} \frac{\partial \by}{\partial \bX} \right|  \\
			&= \left|\frac{\partial \bx_+ }{\partial \bX} \right| Tr\left[ \left(\frac{\partial \bx_+}{\partial \bX}\right)^{-1} \frac{\partial \by}{\partial \bX} \right],
			\end{aligned}
		\end{equation}
		where $Tr(\cdot)$ denotes the trace of a matrix. Thus, we have the conservative force corresponding to the entropy contribution:
		\begin{equation}\label{Fcon+ent}
			 \bm{F}_{con,+}^{ent}  = -\nabla P_+ - \mu_0 \nabla  \rho_-,
		\end{equation}
		where $$\displaystyle \mu_0 = \frac{\partial \Psi_0}{\partial \rho_+} =\frac{\partial \Psi_0}{\partial \rho_-} = -k_BT \left[ \log(1 - v\rho_+ -v\rho_-) + 1\right], $$
		 and the thermal pressure 
		\[
			\begin{aligned}
		   \displaystyle P_+ &= \Psi_+ + \Psi_0 -\frac{\partial (\Psi_+ +\Psi_0)}{\partial \rho_+}  \frac{\rho_+}{J_+} \\
		                               &= k_B T ( \rho_- -\frac{1}{v} ) \log\left( 1-v\rho_+ - v\rho_-\right). 
		 	\end{aligned}
		\]
		The conservative force~\reff{Fcon+ent} can be further expressed as
		\[
		\begin{aligned}
			 \bm{F}_{con,+}^{ent} &= -\left(k_BT\nabla \rho_+ + k_BT\frac{v\rho_+}{1-\sum\limits_{i = \pm}v\rho_i}\sum_{j = \pm}\nabla \rho_j  +\frac{\partial P_+}{\partial T}\nabla T\right)\\
			&=k_B (\frac{1}{v} - \rho_- ) \log(1 - v\rho_+ -v\rho_-) \nabla T  - k_B T  \frac{  v\rho_+ \nabla \rho_- + (1 - v \rho_-) \nabla \rho_+ } { 1 - v\rho_+ -v \rho_-}. 
		\end{aligned}
		\]
		Along with $\bm{F}_{con,+}^{pot} $ given by ~\reff{F^pot_con+} in the main text, we arrive at the total conservative force \reff{con} for cations in the main text. 
		
		The conservative force corresponding to the entropy contribution for anions  can be obtained by taking the variation of the action $A_{ent}$ with respect to the flow map $\bx_{-}(\bX ,t)$:
		\[
		\begin{aligned}
			\bm{F}_{con,-}^{ent} &= -\nabla P_- - \mu_0 \nabla  \rho_+\nonumber\\
			&= -\left(k_BT\nabla \rho_- + k_BT\frac{v\rho_-}{1-\sum\limits_{i = \pm}v\rho_i}\sum_{j = \pm}\nabla \rho_j  +\frac{\partial P_-}{\partial T}\nabla T\right)\\
			&= k_B (\frac{1}{v} - \rho_+ ) \log(1 - v\rho_+ -v\rho_-) \nabla T  - k_B T  \frac{  (1 - v\rho_+)\nabla \rho_- + v\rho_- \nabla \rho_+ } { 1 - v\rho_+ -v\rho_-},
		\end{aligned}
		\]
		where, analogous to the derivation for cations, the thermal pressure due to anions is defined by
		\[
		\begin{aligned}
		\displaystyle P_- &= \Psi_- + \Psi_0 -\frac{\partial (\Psi_- +\Psi_0)}{\partial \rho_-}  \frac{\rho_-}{J_-} \\
		                           &= k_B T (\rho_+ -\frac{1}{v} ) \log(1 - v\rho_+ -v\rho_-).
		\end{aligned}
		\]
		With $\bm{F}_{con,-}^{pot} = -q_-\rho_{-}\nabla\phi$ that is derived analogously to~\reff{F^pot_con+} in the main text, we arrive at the total conservative force \reff{con} for anions.

		\subsection{\label{sec:level2} Derivation of temperature equation}
		We use the first and second laws of thermodynamics to derive the equation for temperature~\reff{th1eq} in the main text.
		The time derivative of the internal energy reads
		\begin{equation}
			\begin{aligned}
				\frac{d}{dt} U(V,t) &= \frac{d}{dt} \int_{V} \left( \Psi\left( \rho_+(\bx, t), \rho_-(\bx, t), T(\bx,t)\right) - T(\bx, t)\frac{\partial \Psi\left( \rho_+(\bx, t), \rho_-(\bx, t), T(\bx,t)\right) }{\partial T}  \right)d\bx \\
				& \qquad+ \frac{d}{dt} \left[ \sum_{i,m = \pm}\frac{q_iq_m}{2}\iint_{V}\rho_i(\bx,t)\rho_m(\bx^{\prime},t)G(\bx,\bx^{\prime}) d\bx^{\prime} d\bx  \right. \\   
				& \left.  \qquad \qquad \quad+\sum_{i= \pm} q_i\int_{V}\rho_i(\bx)\left( \psi_X(\bx,t) + \sum_{m = \pm}\int_{\Omega \backslash V}q_m\rho_m(\bx^{\prime}, t)G(\bx,\bx^{\prime}) d\bx^{\prime} \right)   d\bx \right] \\
				& =  \int_{V}\frac{k_BC_0}{v}\frac{\partial T}{\partial t}d\bx + \sum_{i = \pm}q_i\int_{V}\frac{\partial \rho_i}{\partial t}\left( \psi_X(\bx,t) + \sum_{m=\pm}\int_{\Omega}q_m\rho_mG(\bx,\bx^{\prime})d\bx^{\prime}  \right) d\bx   \\
				& \qquad +\sum_{i = \pm}q_i\int_{V}\rho_i\frac{\partial}{\partial t}\left[ \psi_X(\bx,t) + \sum_{m = \pm}\int_{\Omega \backslash V}q_m\rho_m(\bx^{\prime}, t)G(\bx,\bx^{\prime}) d\bx^{\prime} \right] d\bx \\
				& = \int_{V}\left[ \frac{k_BC_0}{v}\frac{\partial T}{\partial t}  +\sum_{i =\pm}q_i\frac{\partial \rho_i}{\partial t} \phi\right]d\bx + \frac{dW}{dt} +\sum_{i = \pm}\int_{V}\nabla\cdot(P_i\bu_i)d\bx  \\
				& = \int_{V}\frac{k_BC_0}{v}\frac{\partial T}{\partial t}d\bx -J_{E}  + \frac{dW}{dt} +\sum_{i = \pm}\int_{V}\nabla\cdot(P_i\bu_i)d\bx + \int_{V}\sum_{i = \pm}q_i\rho_i\bu_i \cdot \nabla \phi d\bx,
			\end{aligned}
		\end{equation}
		where~\reff{dW}, \reff{JE}, and mass conservation~\reff{masscon} in the main text have been used in the calculation. 
		Combining with the energy conservative law \reff{th1law},  we have by the arbitrariness of the control volume $V$ that
		\begin{equation}
			\label{th1eqo}
			\frac{k_BC_0}{v}\frac{\partial T}{\partial t} = -\nabla\cdot \bm j_h -\sum_{i = \pm}\nabla\cdot(P_i\bu_i) - \sum_{i = \pm}q_i\rho_i\bu_i\cdot\nabla\phi.
		\end{equation}
		
		On the other hand, the second law of thermodynamics reads
		\begin{align}
			\label{secTH}
			\frac{dS(V,t)}{dt} +\int_{\partial V}\frac{\bm j_h}{T}\cdot d\bS +\bm J_s = \Delta(V,t),
		\end{align}
		where $\bm{j}_h$ is for the heat flux, $\Delta(V,t)$ is the dissipation functional,  and $\bm J_s$ is the entropic flux  at boundary $\partial V$ given by
		\begin{align}
			\bm J_s = -\sum_{i = \pm}\int_{\partial V}\frac{\partial \left( \Psi_i + \Psi_0\right) }{\partial T}\bu_i\cdot d\bS.
		\end{align}
		The time derivative of the entropy \reff{S} reads
		\[
		\begin{aligned}
			\frac{d}{dt} S(V,t) &= \int_{V} \left[  \frac{k_BC_0}{vT}\frac{\partial T}{\partial t} - \sum_{i = \pm}\frac{\partial^2\Psi_i}{\partial T\partial\rho_i}\frac{\partial \rho_i}{\partial t}- \sum_{i = \pm}\frac{\partial^2\Psi_0}{\partial T\partial\rho_i}\frac{\partial\rho_i}{\partial t}\right] d\bx   \\
			&= \int_{V} \left[  \frac{k_BC_0}{vT}\frac{\partial T}{\partial t} + \sum_{i = \pm}\frac{\partial^2\Psi_i}{\partial T\partial\rho_i}\nabla\cdot(\rho_i\bu_i)+  \sum_{i = \pm}\frac{\partial^2\Psi_0}{\partial T\partial\rho_i}\nabla\cdot(\rho_i\bu_i)\right] d\bx   \\
			& = \int_{V} \left[  \frac{k_BC_0}{vT}\frac{\partial T}{\partial t}  +\sum_{i = \pm}\frac{\partial^2 \Psi_i}{\partial T\partial\rho_i}\left( \rho_i\nabla\cdot\bu_i + \bu_i\cdot\nabla\rho_{i}\right) + \sum_{i = \pm}\frac{\partial^2\Psi_0}{\partial T\partial\rho_i}\left( \rho_i\nabla\cdot\bu_i + \bu_i\cdot\nabla\rho_{i}\right) \right] d\bx   \\
			& = \int_{V} \left[  \frac{k_BC_0}{vT}\frac{\partial T}{\partial t}  +\sum_{i = \pm}\left( \frac{\partial^2 \Psi_i}{\partial T\partial\rho_i}\rho_i\nabla\cdot\bu_i + \nabla\cdot\left( \frac{\partial\Psi_i}{\partial T}\bu_i\right) -\frac{\partial \Psi_i}{\partial T}\nabla\cdot\bu_i\right)  \right. \\
			& \qquad + \left. \sum_{i = \pm}\left(  \frac{\partial^2 \Psi_0}{\partial T\partial\rho_i}\rho_i\nabla\cdot\bu_i  +  \nabla \frac{\partial \Psi_0}{\partial T}  \right) - \bu_-\frac{\partial^2\Psi_0}{\partial T\partial\rho_+} \nabla\rho_+  - \bu_+\frac{\partial^2\Psi_0}{\partial T\partial\rho_-} \nabla\rho_-   \right] d\bx \\
			& = \int_{V} \left[  \frac{k_BC_0}{vT}\frac{\partial T}{\partial t}  +\sum_{i = \pm}\left( \frac{\partial^2 \Psi_i}{\partial T\partial\rho_i}\rho_i\nabla\cdot\bu_i + \nabla\cdot\left( \frac{\partial\Psi_i}{\partial T}\bu_i\right) -\frac{\partial \Psi_i}{\partial T}\nabla\cdot\bu_i\right)  \right. \\
			& \qquad \quad+ \left.  \sum_{i = \pm} \left( \left( \rho_i\frac{\partial^2\Psi_0}{\partial T\partial\rho_i}-\frac{\partial\Psi_0}{\partial T}\right)\nabla\cdot\bu_i + \nabla\cdot\left(\frac{\partial \Psi_0}{\partial T}\bu_i \right) \right) \right.\\
			& \qquad \quad- \left. \bu_-\frac{\partial^2\Psi_0}{\partial T\partial\rho_+} \nabla\rho_+  - \bu_+  \frac{\partial^2\Psi_0}{\partial T\partial\rho_-} \nabla\rho_-    \right] d\bx   \\
			& = \bm J_s + \int_{V} \left[  \frac{k_BC_0}{vT}\frac{\partial T}{\partial t} + \sum_{i = \pm} \left( \rho_i\frac{\partial^2(\Psi_i+ \Psi_0)}{\partial T\partial\rho_i}-\frac{\partial(\Psi_i+\Psi_0)}{\partial T}\right)\nabla\cdot\bu_i \right.\\
			& \qquad - \left. \bu_-\frac{\partial^2\Psi_0}{\partial T\partial\rho_+} \nabla\rho_+  - \bu_+  \frac{\partial^2\Psi_0}{\partial T\partial\rho_-} \nabla\rho_-    \right] d \bx \\
			& = \bm J_s + \int_{V} \left[  \frac{k_BC_0}{vT}\frac{\partial T}{\partial t} + \sum_{i = \pm} \left( \frac{\partial P_i}{\partial T}\nabla\cdot\bu_i\right) - \bu_-\frac{\partial^2\Psi_0}{\partial T\partial\rho_+} \nabla\rho_+  - \bu_+  \frac{\partial^2\Psi_0}{\partial T\partial\rho_-} \nabla\rho_-  \right] d \bx. \\
		\end{aligned}
		\]
		Combining with the second law of thermodynamics,  we have by the arbitrariness of the control volume $V$ that
		\begin{equation}
			\label{th2eqo}
			\frac{k_BC_0}{vT}\frac{\partial T}{\partial t} + \sum_{i = \pm} \left( \frac{\partial P_i}{\partial T}\nabla\cdot\bu_i\right) -\bu_-\frac{\partial^2\Psi_0}{\partial T\partial\rho_+} \nabla\rho_+  - \bu_+  \frac{\partial^2\Psi_0}{\partial T\partial\rho_-} \nabla\rho_-    = -\nabla\cdot\left( \frac{\bm j_h}{T}\right) +\widetilde{\Delta},
		\end{equation}
		where $\widetilde{\Delta}$ is entropy production density.   Multiplying the equation \reff{th2eqo} by the local temperature $T(\bx,t)$ and then  subtracting from the equation~\reff{th1eqo} give
		\begin{equation}
			\label{forcebb}
			\sum_{i = \pm}\left( \bm F_{con, i} - \bm F_{dis, i}\right)\cdot\bu_i = \bm j_h\cdot\left( \frac{\bm j_h}{k_BT} +\frac{\nabla T}{T}\right), 
		\end{equation}
		where $\bm F_{con, i} $ and $\bm F_{dis, i}$ are given in \reff{con} and \reff{diss} in the main text.   By the force balance, we have $\bm F_{i}^{con} = \bm F_{i}^{dis}$ and  the heat flux $\bm j_h = -k_B\nabla T$, which is exactly the Fourier law. Finally, substituting the flux into the equation~\reff{th1eqo} leads to the temperature equation~\reff{th1eq} in the main text.
		
		\subsection{Numerical methods}
		The governing system~\reff{PNPFS-dimensionless} with initial and boundary conditions is integrated first in time with a fully implicit scheme~\reff{PNPFS-timediscretization}. It then becomes a boundary value problem (BVP) with very thin boundary layers, which is further solved with the BVP4C software in the MATLAB~\cite{Kierzenka01, Kierzenka08}. 
		To use the BVP4C software, we reformulate the BVP~\reff{PNPFS-timediscretization} into the following first-order differential equations by introducing additional variables:
		\[
		\left\{
		\begin{aligned}
			& \frac{\partial}{\partial x}\phi^{n+1} = \mu^{n+1}, \\
			& \frac{\partial}{\partial x}\mu^{n+1}  = -\frac{\rho^{n+1}}{\epsilon^2},  \\
			& \frac{\partial}{\partial x}c^{n+1} = \frac{vc^{n+1}-1}{T^{n+1}}\left( j_c^{n+1}+\rho^{n+1} \mu^{n+1} -\frac{2-vc^{n+1}}{v}s^{n+1}W^{n+1} \right) , \\
			& \frac{\partial}{\partial x}j_c^{n+1}  = \frac{c^{n}-c^{n+1}}{\epsilon \Delta t },\\
			& \frac{\partial}{\partial x}\rho^{n+1} = \frac{1}{T^{n+1}}\left[ -j_{\rho}^{n+1}-c^{n+1}\mu^{n+1}+v\rho^{n+1}(j_c^{n+1}+\rho^{n+1}\mu^{n+1}+\frac{vc^{n+1}-1}{v}s^{n+1} W^{n+1})\right] ,\\
			&\frac{\partial}{\partial x}j_{\rho}^{n+1} = \frac{\rho^{n}-\rho^{n+1}}{\epsilon \Delta t},\\
			&\frac{\partial}{\partial x}T^{n+1} = W^{n+1},\\
			&\frac{\partial}{\partial x}W^{n+1} = \frac{1}{k}\left( \frac{C_0}{v}\frac{T^{n+1}-T^{n}}{\Delta t} + \epsilon\frac{\partial}{\partial x}\left(  \sum_{i = \pm}P_i^{n+1} u_i^{n+1}\right) + \epsilon j_{\rho}^{n+1}\mu^{n+1}\right),
		\end{aligned}
		\right.
		\]
		where $s^{n+1} = \log(1-vc^{n+1})$, $P_{\pm}^{n+1}=\left( c^{n+1} \mp \rho^{n+1} -\frac{1}{v}\right) T^{n+1}s^{n+1} $, and  $u_{\pm}^{n+1}= \frac{j_c^{n+1} \pm j_{\rho}^{n+1}  }{c^{n+1} \pm \rho^{n+1} }$. With additional variables, the boundary conditions are given by
		\begin{equation}
			\label{PNPFS_bvpbc}
			\left\{
			\begin{aligned}
				\phi^{n+1}(\pm L) \pm H\mu^{n+1} (\pm L) &= \mp\phi_s, \\
				j_c^{n+1}(\pm L) &= 0, \\
				j_{\rho}^{n+1}(\pm L) &= 0, \\
				W^{n+1}(\pm L) &= 0,
			\end{aligned}
			\right.
		\end{equation}
		where $H$ represents the width of Stern layers.  Notice that the Stern/diffuse layer interface is located at $x = \pm L$; cf. Fig.~\ref{f:geometry} in the main text. Here $H$ and $L$ have been nondimensionalized with a reference length scale.  Extensive numerical results evidence that the numerical methods can accurately and robustly resolve the thin boundary layers.

		
%

\end{document}